\documentclass[journal]{IEEEtran}
\usepackage[utf8]{inputenc}
\usepackage[english]{babel}
\usepackage{amsmath}
\usepackage{mathtools,amssymb, nccmath}
\usepackage{amsfonts}
\usepackage{amssymb}
\usepackage{amsthm}
\usepackage{amsmath,amssymb,amsfonts}
\usepackage{algorithmic}
\usepackage{graphicx}
\usepackage{textcomp}
\usepackage{xcolor}
\usepackage{graphics}
\usepackage{graphicx}
\usepackage{epsfig}
\usepackage{fancyhdr}
\usepackage{enumitem}
\usepackage{wasysym}
\usepackage{latexsym}
\usepackage{color}
\usepackage{subfig}
\usepackage{cite}
\usepackage{mathtools}
\usepackage[linesnumbered,ruled,vlined]{algorithm2e}
\usepackage[left=2cm,right=2cm,top=2cm,bottom=2cm]{geometry}
\newcommand{\norm}[1]{\left\lVert#1\right\rVert}
\begin{document}
\title{Cooperative Hybrid Transmit Beamforming in Cell-free mmWave MIMO Networks}
%\author{\IEEEauthorblockN{Meesam Jafri},
%\and
%\IEEEauthorblockN{ Suraj Srivastava},
%\and
%and
%\IEEEauthorblockN{Aditya K. Jagannatham}
%\and
%}
\author{ Meesam~Jafri,\IEEEmembership{ Student Member,~IEEE,}  Suraj~Srivastava,\IEEEmembership{ Student Member, IEEE,} Naveen~K.~D.~Venkategowda,\IEEEmembership{ Student Member, IEEE,} Aditya~K.~Jagannatham, \IEEEmembership{ Member, IEEE,} Lajos~Hanzo, \IEEEmembership{ Life Fellow, IEEE \vspace{-20pt}}
\thanks{Copyright (c) 2015 IEEE. Personal use of this material is permitted.
However, permission to use this material for any other purposes must be obtained
from the IEEE by sending a request to pubs-permissions@ieee.org. }
\thanks{The work of Aditya K. Jagannatham was supported in part by the Qualcomm Innovation Fellowship, and in part by the Arun Kumar Chair Professorship.}
\thanks{L. Hanzo would like to acknowledge the financial support of the Engineering and Physical Sciences Research Council projects EP/W016605/1 and EP/P003990/1 (COALESCE) as well as of the European Research Council's Advanced Fellow Grant QuantCom (Grant No. 789028)}
\thanks{M. Jafri, S. Srivastava and A. K. Jagannatham are with the
Department of Electrical Engineering, Indian Institute of Technology,
Kanpur, Kanpur, 208016, India (e-mail: meesam@iitk.ac.in, ssrivast@iitk.ac.in
adityaj@iitk.ac.in.)}

\thanks{N. K. D. Venkategowda is with the Department of Science and Technology, Link\"oping University, 60174 Norrk\"oping, Sweden (e-mail: naveen.venkategowda@liu.se)}

\thanks{L. Hanzo is with the School of Electronics and Computer
Science, University of Southampton, Southampton SO17 1BJ, U.K.
(e-mail: lh@ecs.soton.ac.uk)}
}
%\thanks{L. Hanzo is with the School of Electronics and Computer
%Science, University of Southampton, Southampton SO17 1BJ, U.K.
%(e-mail: lh@ecs.soton.ac.uk)}
\maketitle
\begin{abstract}
Hybrid precoders and combiners are designed for cooperative cell-free multi-user millimeter wave (mmWave) multiple-input multiple-output (MIMO) cellular networks for low complexity interference mitigation. Initially, we derive an optimal hybrid transmit beamformer (HTBF) for a broadcast scenario considering both total and per access point (AP) power constraints. Next, an optimal successive hybrid beamformer technique is proposed for unicast and multicast scenarios which relies on the optimal minimum variance distortionless response (MVDR). We demonstrate that it mitigates both the interuser and intergroup interference, while successively ensuring orthogonality to the previously scheduled users/user groups. Furthermore, it is shown theoretically that the proposed schemes are capable of supporting a large number of users. Subsequently, a Bayesian learning (BL) based method is conceived for jointly designing the RF and baseband precoders/combiners for the various scenarios considered. Furthermore, we also conceive the uplink counterpart of our HTBF scheme, which is based on maximizing the signal-to-interference-plus noise ratio (SINR) of each individual user. Finally, the efficacy of the proposed schemes is characterized by our extensive simulation results in terms of cancelling the interuser/intergroup interference, which improves the spectral efficiency. 
%Simulation results are presented to demonstrate the superior capacity performance of the proposed multi-cell cooperative beamforming schemes.
%Base stations (BSs) are expected to be densely deployed in 5G mmWave MIMO wireless networks, such systems are going to be greatly suffers from the both intra-cell and inter-cell interference (ICI). To mitigate the interferences, cooperative beamformer design is the efficient solution, wherein different BSs cooperate to jointly design the beamformer. In this work, we derive an hybrid broadcast beamforming (HBB) scheme for multi-cell cooperative broadcast transmission for a pooled power constraint at the base stations with limited number of RF chains. This scheme is extended to include per base station power constraints by formulating it as a relaxed semi-definite program (SDP).
%Next, we consider unicast/multicast mmWave multi-cell cooperative transmission with interference among users/user group. In this context, this paper proposes the optimal minimum variance distortionless response (MVDR)- based successive minimum variance hybrid beamforming (OSHB) scheme that effectively mitigate the ICI. We also consider the multi-cell cooperative uplink scenario and derive a successive multiuser (MU) uplink beamforming (SCUHB) scheme that maximizes the signal-to-interference-plus noise ratio (SINR) of each user. Simulation results are presented to demonstrate the superior capacity performance of the proposed multi-cell transmission schemes.
\end{abstract}
\section{Introduction}
The next-generation wireless communication systems aim for increased data rates, reduced lower latency and large-scale access. The mmWave technology has shown significant promise in meeting these requirements \cite{hemadeh2017millimeter, 8424038, shokri2015millimeter}. Communication at mmWave frequencies has recently attracted significant attention due to the fact that the sub-6 GHz band faces acute spectrum shortage \cite{8565897, 8770151}. In this context, the under-utilized mmWave band has the potential of satisfying the requirements of next generation wireless networks. Furthermore, the abundance of free spectrum in the mmWave band is capable of supporting the massive connectivity of devices required by the Internet of Things (IoT) \cite{sahoo2018enabling}.
However, as shown via practical measurements in \cite{heath2016overview}, \cite{sohrabi2016hybrid}, mmWave networks suffer from signal blockage and penetration losses, which degrades the received signal strength. 
The dense deployment of access points (APs) for creating small cells has shown significant promise for overcoming these challenges. However, inter-cell interference (ICI) becomes a significant bottleneck with the ever-increasing network densification. Several cooperative signal processing concepts, such as cooperative MIMO networks, network MIMO, and coordinated multipoint (CoMP) transmission \cite{gesbert2010multi, venkatesan2007network, irmer2011coordinated}, are presented in the literature to address the issue of ICI. Note that the network architectures mentioned above are conventionally implemented in a network-centric fashion by dividing the APs into disjoint clusters. In \cite{ngo2017cell, nayebi2017precoding, ammar2021user}, a user-centric cell-free MIMO structure relying on a large number of distributed APs is considered, which are connected to a central processing unit (CPU). The network jointly serves multiple users in conjunction with no cell boundaries, similar to the cooperative MIMO structure. However, cooperative MIMO networks mitigate the ICI by grouping multiple cells into fixed cooperative clusters, which shifts the interference management from the cell level to the cluster level. The resultant inter-cluster interference imposes a fundamental limit on cooperation. Hence, users close to the cluster edge might not benefit from the cooperative MIMO clustering. By contrast, cell-free MIMO facilitates improved service for cell-edge users due to the elimination of the inter-cluster interference. Furthermore, the other significant aspect of cell-free MIMO is the deployment of a higher number of APs than that of the users in comparison to the cooperative MIMO. Thus, cell-free networks promise a further improvement in spectral efficiency by mitigating the interference through intelligent, cooperative beamforming.

Apart from the benefit of wider bandwidth availability, the short wavelength of mmWave frequencies allows us to pack large antenna array into a compact space for supporting highly directional beamforming, while simultaneously serving numerous users. Thus, the multiple-input-multiple-output (MIMO) technology can be exploited to obtain high beamforming gains for mitigating the effects of path loss, atmospheric absorption and penetration loss in the mmWave regime \cite{rangan2014millimeter}.
However, the high power consumption and associated cost/complexity of assigning a dedicated per-antenna RF chain prohibits the employment of the conventional fully-digital (FD) transceiver architecture in mmWave MIMO systems \cite{heath2016overview}.
This has motivated the conception of the novel hybrid MIMO transceiver architecture as a promising solution, which employs much fewer RF chains in comparison to the antenna elements, thus easing its practical implementation. The overall MIMO signal processing in such a hybrid architecture is divided into the RF and baseband domains, where RF transmit precoding (TPC) yields a directional beamforming gain, whereas spatial multiplexing and interference cancellation are achieved via suitable baseband TPC. It can be readily observed that the design of the baseband and RF precoders and combiners is the key to address the various challenges associated with the proposed cooperative cell-free multi-user (CFMU) mmWave MIMO system, and also to harness the significant gains promised by the mmWave band for 5G/B5G. 
\subsection{Related Works}
In the recent on mmWave literature, sophisticated techniques have been proposed for designing hybrid TPCS and receive combiners (RCs) for mmWave MIMO systems \cite{el2014spatially, wang2019low, noh2016zero, alkhateeb2015limited}. In particular, the early contributions focused more on single-user (SU) scenarios.
Omar \textit{et al.,} in \cite{el2014spatially} considered SU mmWave hybrid MIMO systems and derived the hybrid transmit beamformers (HTBF) in which the sparse scattering nature of mmWave wireless channel has been exploited to obtain a two-stage solution via the popular orthogonal matching pursuit (OMP) algorithm. Wang \textit{et al.,} \cite{wang2019low}, proposed another hybrid TPC design that decomposes the original HTBF design problem into several sub-problems to render the optimization tractable. 
In order to further improve the spectral efficiency and to mitigate the inter-user interference (IUI) in practical mmWave cellular systems a multi-user (MU) mmWave MIMO system has also been considered in the pioneering works \cite{noh2016zero}, \cite{alkhateeb2015limited}, wherein a BS simultaneously serves multiple users/devices. 

In a mmWave MU MIMO system, the digital component of the hybrid architecture provides higher degrees of freedom than analog-only beamforming, and can hence be employed for mitigating the interference among the different users. Note that the design of HTBF is a key challenge in MU mmWave MIMO systems, since the resultant IUI degrades the overall system performance \cite{ding2015performance}, \cite{zhang2015downlink}. Therefore, there is an urgent need for minimizing the IUI, while the HTBF gain has to be simultaneously maximized for each user. Thus, it is of pivotal importance to design efficient HTBF for mmWave MU MIMO systems that have a low complexity. Hence, sophisticated HTBF techniques have been developed for optimizing their performance. In \cite{alkhateeb2015limited}, a low-complexity iterative algorithm was proposed, which exploited the sparse scattering nature of the mmWave wireless channel for designing the HTBF of the BS and analog RC of each user. The proposed solution was shown to be asymptotically optimal for the large-scale antenna regime, which also demonstrates the well-known fact that MU MIMO systems attain a high capacity than SU MIMO systems.

In \cite{hong2013joint, zhang2010cooperative, li2015joint, cheng2013joint, yang2017opportunistic}, the coordination of different base stations (BSs) has been proposed for conventional sub-6 GHz systems for improving their performance. However, the proposed schemes targeted exclusively fully-digital sub-6 GHz MU MIMO systems. Hence, they are unsuitable for mmWave MU MIMO cellular systems.
Michaloliakos \textit{et al.} \cite{michaloliakos2016joint}, proposed a novel scheme relying on jointly selecting the analog TPCS for maximizing the data rate of the users by harnessing a set of predefined beam patterns. Zhu \textit{et al.} \cite{zhu2017hybrid}, proposed a novel cooperative HTBF design for multi-cell multi-user (MCMU) scenarios based on the Kronecker decomposition. First, analog transmit beamformer (TBF) is obtained by decomposing the beamformer as well as the data and interference
path-vectors into the corresponding Kronecker products of unit-modulus phase-shift vectors. Next, the baseband TBF is designed by using the minimum mean square error (MMSE) TBF technique. Shu \textit{et al.} \cite{sun2018analytical}, presented both signal-to-leakage-plus-noise-ratio (SLNR) and regularized zero-forcing based HTBF techniques for multi-stream MCMU systems to study the benefits of interference coordination in terms of the spectral efficiency.
Bai \textit{et al.} \cite{bai2018cooperative}, presented a two-step hybrid cooperative TPC design for a distributed antenna scenario combined with user selection for energy-efficient transmissions. While their work has immense utility, it is restricted to a single-cell scenario, which does not exploit the additional advantages offered by multi-cell cooperation. Fang \textit{et al.} \cite{fang2021hybrid}, maximize the sum-power consumption of the APs supporting single-antenna users relying on cooperative HTBF. 

While several authors have addressed the problem of hybrid beamformer design for classic cellular systems, the research of their cell-free counterparts is limited. In \cite{kim2019joint}, the authors proposed a hybrid beamformer scheme, where the analog and digital beamformers are alternately optimized based on the weighted minimum mean squared error (WMMSE) criterion. In \cite{hou2017joint}, the authors proposed a two-step procedure for their hybrid beamformer design. In the first step, the analog TPC is obtained from a codebook, followed by semidefinite relaxation (SDR) and convex approximation-aided baseband TPC design. In \cite{femenias2019cell}, the authors designed a hybrid beamforming scheme for CFMU massive MIMO systems toward maximizing the user rate. Chenghao \textit{et. al.} \cite{feng2021weighted} proposed a block coordinated descent (BCD) algorithm based hybrid transmit beamformer (HTBF) maximizing the weighted sum-rate of cell-free mmWave MIMO systems. However, the aforementioned treatises assume the availability of either perfect knowledge of CSI or that of its second-order statistics. The HTBF design of a cell-free network requires the full knowledge of AoAs/AoDs of the multipath components and their complex-valued path gains. However, in practice, acquiring accurate estimates of these quantities is challenging. In the recent studies of cell-free MIMO systems, sophisticated techniques have been proposed for channel estimation in CFMU mmWave MIMO systems \cite{jin2019channel, souza2021effective, guo2021joint, song2021joint, wang2022two}. Furthermore, the channel matrix in the mmWave regime is sparse, which can be jointly attributed to the reduced scattering and diffraction effects as well as to the focused beam emanating from the large antenna arrays. Exploiting this sparsity can lead to a significant reduction in the number of pilot transmissions required and also a substantial improvement in the estimation accuracy. In this context, the compressed sensing (CS) techniques proposed in \cite{gilbert2005applications, barbotin2012estimation} utilize a sparse signal recovery model for channel estimation. The authors of \cite{huang2018iterative, srivastava2021sparse} have demonstrated the applicability of some classical CS based schemes for sparse channel estimation at a reduced pilot overhead. One can readily employ these existing schemes for efficient sparse channel estimation in CFMU mmWave MIMO networks.

In wireless communication systems, there are several scenarios, where common data has to be sent to a group of users, for instance, during the transmission of content pertaining to a live event. In such scenarios, multicast is an efficient framework for improving both the spectral and power efficiency by simultaneously transmitting the common data to a group of users \cite{dartmann2013low, guo2018multi, dong2020multi}. This has huge potential in numerous applications, such as mobile TV, live video streaming, video conferencing etc.
As a further advance in \cite{dartmann2013low}, a low-complexity uplink-downlink duality based suboptimal algorithm is presented to jointly design the beamformers for multiple users in a single-cell wireless multicast network. 
Chengjun \textit{et al.} \cite{guo2018multi} conceived a low-complexity multi-quality multicast beamforming schemes..
The authors of \cite{dong2020multi} proposed a weighted MMSE based multicast beamformer employing the popular successive convex approximation and Lagrangian duality based techniques.  
Note that all the above contributions only consider conventional sub-6 GHz fully digital transceiver architectures, which cannot be used in mmWave CFMU MIMO systems. Multicast transmission,  specifically designed for mmWave MIMO networks, was studied in \cite{choi2015iterative,dai2015hybrid,wang2018hybrid}.
In \cite{choi2015iterative}, a single group scenario is considered for HTBF-aided multicast transmission, wherein an iterative algorithm was designed based on the principle of alternating minimization. In the seminal work \cite{dai2015hybrid}, the author presented a single-cell single-group HTBF design relying on max-min fairness for multicasting using a limited number of RF chains. In \cite{wang2018hybrid}, a suboptimal low-complexity HTBF design was proposed for the downlink of unicast and multicast mmWave MIMO systems. However, all these papers only considered a single AP scenario for mutlicast transmission. To the best of our knowledge, none of the works in the existing literature have sophistically addressed the problem of cooperative HTBF design in CFMU mmWave MIMO systems. Thus, motivated by these limitations of the existing literature reviewed above, this paper presents a novel cooperative beamformer design conceived for efficiently mitigating the interference in CFMU mmWave MIMO networks. Table \ref{tab_lit_rev} boldly and explicitly contrasts to the literature discussed above. Next, we further detail the various contributions of our work.
\begin{table*}

    \centering

    \caption{Contrasting our contribution to the existing literature} \label{tab_lit_rev}

\begin{tabular}{|l|c|c|c|c|c|c|c|c|c|c|c|c|c|}
    \hline

 & \cite{el2014spatially} &  \cite{wang2019low} & \cite{alkhateeb2015limited} & \cite{hong2013joint} & \cite{zhu2017hybrid} & \cite{sun2018analytical} & \cite{femenias2019cell} & \cite{feng2021weighted} & \cite{dartmann2013low} & \cite{choi2015iterative} & \cite{dai2015hybrid} & \cite{wang2018hybrid} & Proposed Work \\ [0.5ex]

\hline
mmWave communication & \checkmark & \checkmark & \checkmark &  & \checkmark & \checkmark & \checkmark & \checkmark &  & \checkmark & \checkmark & \checkmark & \checkmark\\
\hline
Hybrid architecture & \checkmark & \checkmark & \checkmark &  & \checkmark & \checkmark & \checkmark & \checkmark &  & \checkmark & \checkmark & \checkmark & \checkmark\\
\hline
Multi-user &  &  & \checkmark & \checkmark & \checkmark & \checkmark & \checkmark & \checkmark & \checkmark & \checkmark & \checkmark & \checkmark & \checkmark \\
\hline
Downlink & \checkmark & \checkmark & \checkmark & \checkmark &  & \checkmark & \checkmark & \checkmark & \checkmark & \checkmark & \checkmark & \checkmark &  \checkmark \\
\hline
Uplink &  &  &  &  & \checkmark &  &  &  & \checkmark & & & &  \checkmark \\
\hline
Cooperative beamforming &  &  &  & \checkmark & \checkmark & \checkmark & \checkmark & \checkmark & \checkmark & & & & \checkmark\\
\hline
Multicast single-group &  &  &  &  &  &  &  &  & \checkmark & \checkmark & \checkmark & \checkmark & \checkmark\\
\hline
Multicast multi-group  &  &  &  &  &  &   &  &  & \checkmark & &  &  & \checkmark\\
\hline
Broadcast &  &  &  &  &  &  &  &  &  & \checkmark & & \checkmark & \checkmark\\
\hline
Unicast  &  &  &  & \checkmark & \checkmark & \checkmark  & \checkmark & \checkmark & \checkmark & & & \checkmark & \checkmark\\
\hline
%Robust design &  &  &  &  &  &   &  &  &  & & \checkmark\\
%\hline
\end{tabular}
\end{table*}
\subsection{Contributions}
Again, we design HTBFs and RCs for broadcast, unicast and multicast scenarios in cooperative CFMU mmWave MIMO cellular systems.
\begin{itemize}
\item Initially, we design an optimal hybrid broadcast beamformer for CFMU mmWave MIMO systems. Furthermore, the optimal TBFs having per-AP power constraints are also derived for this scenario. At the users, a low-complexity maximal ratio combiner (MRC) is used in the baseband for maximizing the effective SNR.
\item Next, the design and analysis is extended to both unicast and multicast transmission scenarios. Toward this, the minimum variance distortionless response (MVDR) \cite{johnson1992array}, \cite{trees2002optimum} based optimal successive minimum variance hybrid beamforming (OSHB) schemes are designed for mitigating both the inter-user and inter-group interference. It has been demonstrated theoretically that the proposed schemes can support a higher number of users than popular block diagonalization (BD) method.
\item Subsequently, a Bayesian learning (BL)-based method is developed for jointly designing the RF and baseband TPCs/RCs, which is of significant utility, when perfect knowledge of the angles of arrival/departure (AoAs/AoDs) and path gains of the multipath components is not available.
\item Furthermore, in order to address the cooperative beamformer design problem of the uplink, a successive CFMU uplink hybrid beamforming (SCUHBF) scheme is developed that maximizes the SINR. It is shown that the proposed scheme achieves an improved data rate as a benefit of interference cancellation.
\item Simulation results are provided for characterizing the OSHB and SCUHBF schemes in cancelling the interuser/intergroup interference together with the improved spectral efficiency compared to a no-coordination scenario, the performance attained is close to that of the ideal fully-digital design.
\end{itemize}
\subsection{Organization of the paper}
The remainder of the paper is arranged as follows. Section \ref{sec 2} introduces the system model and channel model. In Section \ref{sec HBB}, we formulate the problem of HBBF for CFMU mmWave MIMO systems. Section IV and V describe the problem of cooperative TPC design of unicast and multicast scenarios, respectively, and develop the OSHB schemes for it. The SCUHBF scheme of an uplink scenario is derived in Section VI, followed by the BL-based joint precoder design in Section-VII. Finally, Section VIII presents our simulation results for illustrating the merits of the proposed work, followed by our conclusions in Section IX.
\subsection{Notation}
Vectors and matrices are denoted by boldfaced lower and upper case letters respectively, while scalars are represented by lower case letters. The transpose, Hermitian, conjugate, and inverse of a matrix $\mathbf{F}$ are denoted by $\mathbf{F}^T$, $\mathbf{F}^H$, $\mathbf{F}^*$, and $\mathbf{F}^{-1}$, respectively. Furthermore, $\text{Tr}(\mathbf{A})$ denotes the trace of a matrix $\mathbf{A}$, $\norm{\mathbf{x}}_2$ represents the $l_2$-norm of a vector $\mathbf{x}$, and $\mathop{\mathbb{E}}[\cdot]$ is the expectation operator. Furthermore, ${\mathbf{x}} \sim \mathcal{CN} (\mathbf{b}, \mathbf{A})$ denotes a complex Gaussian random vector having mean $\mathbf{b}$ and covariance matrix $\mathbf{A}$. Finally, $\mathbf{I}_N$ represents an $N \times N$ identity matrix and $\mathbf{A} \succcurlyeq 0$ indicates that $\mathbf{A}$ is a positive semi-definite (PSD) matrix.
\section{mmWave CFMU System and Channel Model}\label{sec 2}
\begin{figure}
\begin{center}
\includegraphics[width = 5cm, height = 4cm]{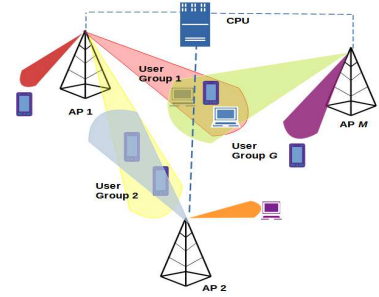}
\caption{Cooperative beamforming in cell-free mmWave MIMO for mulitcast scenario. \vspace{-20pt}}
\label{multicell}
\end{center}
\end{figure}

Consider a downlink cell-free mmWave hybrid MIMO system with $M$ cooperative APs, where the $m$th AP has $N_T$ transmit antennas (TAs) and $N_{\mathrm{RF},m}$ RF chains so that $1 \leq N_{\mathrm{RF},m} << N_T$, $\forall m,  1 \leq m \leq M$. These $M$ APs are connected to a central processing unit (CPU) and serve $G$ user-groups, where the $g$th group is comprised of $U_g$ users, as shown in Fig. \ref{multicell}. Each user is equipped with $N_R$ receive antennas (RAs) and $N_{\mathrm{RF},u}$ RF chains, which satisfy $1 \leq u \leq U_g$, $\forall 1 \leq g \leq G$. We consider a next-generation multicast scenario, where several users request identical services relying on the same data. Such a group of users is referred to as a multicast group in the literature, and many such groups may coexist as shown in Fig. \ref{multicell}. For convenience, user $u$ in the $g$th multicast group is denoted as $U_u^{(g)}$. The signal, $\mathbf{r}^{(g)}_u \in \mathbb{C}^{N_{R} \times 1}$, received at the user $u$ in the $g$th multicast group can be formulated as
\begin{align}
\mathbf{r}^{(g)}_u =& \sum_{m=1}^{M}{\mathbf{H}^{(g)}_{u, m} \mathbf{F}_{\mathrm{RF},m} \mathbf{f}^{(g)}_{\mathrm{BB},m} x^{(g)}} \nonumber \\  +& \sum_{\substack{l=1\\ l \neq g}}^{G} \sum_{m=1}^{M}{\mathbf{H}^{(g)}_{u, m} \mathbf{F}_{\mathrm{RF},m} \mathbf{f}^{(l)}_{\mathrm{BB},m} x^{(l)}} + {\boldsymbol{\delta}}^{(g)}_{u},
\end{align}
where $\mathbf{H}^{(g)}_{u, m} \in \mathbb{C}^{N_{R} \times N_{T}}$ represent the mmWave MIMO channel matrix from the $m$th AP to the user $U_{u}^{(g)}$, $\mathbf{F}_{\mathrm{RF},m} \in \mathbb{C}^{N_{T} \times N_{\mathrm{RF},m}}$ is the RF TPC of the $m$th AP, and $\mathbf{f}^{(g)}_{\mathrm{BB}, m} \in \mathbb{C}^{N_{\mathrm{RF},m} \times 1}$  is the baseband TPC of the $m$th AP for the user-group $g$. Furthermore, ${\boldsymbol{\delta}}^{(g)}_u \in \mathbb{C}^{N_{R} \times 1}$ represents the complex additive Gaussian noise vector with distribution ${\boldsymbol{\delta}}^{(g)} \sim \mathcal{CN} (\mbox{0}, \sigma_{\delta}^2 \mathbf{I}_{N_R})$ and $x^{(g)}$ signifies the transmitted symbol that has unit power i.e., $\mathop{\mathbb{E}}\left[{|x^{(g)}|}^2\right] = \mbox{1}$. The received signal $\tilde{\mathbf{r}}_{u}^{(g)}\in\mathbb{C}^{N_{\mathrm{RF},u}\times 1} $ after RF combining at the user $u$ in the $g$th group is given by
\begin{align*}
\tilde{\mathbf{r}}_{u}^{(g)} =& \left(\mathbf{W}_{\mathrm{RF},u}^{(g)}\right)^H \sum_{m=1}^{M}{\mathbf{H}^{(g)}_{u, m} \mathbf{F}_{\mathrm{RF},m} \mathbf{f}^{(g)}_{\mathrm{BB},m} x^{(g)}}  \nonumber \\ +& \left(\mathbf{W}_{\mathrm{RF},u}^{(g)}\right)^H \sum_{\substack{l=1\\  l \neq g}}^{G} \sum_{m=1}^{M}{\mathbf{H}^{(g)}_{u, m} \mathbf{F}_{\mathrm{RF},m} \mathbf{f}^{(l)}_{\mathrm{BB},m} x^{(l)}} + \tilde{{\boldsymbol{\delta}}}^{(g)}_{u},
\end{align*}
where $\mathbf{W}_{\mathrm{RF},u}^{(g)} \in \mathbb{C}^{N_{R} \times N_{\mathrm{RF}, u}}$ denotes the RF RC of the user $U_{u}^{(g)}$ and $ \tilde{{\boldsymbol{\delta}}}^{(g)}_{u} =  \left(\mathbf{W}_{\mathrm{RF},u}^{(g)}\right)^H{\boldsymbol{\delta}}^{(g)}_{u}$. Let us define the effective channel matrix between the user $U_{u}^{(g)}$ and $m$th AP as 
$\mathbf{H}^{(g)}_{\mathrm{eff},u,m} = \left(\mathbf{W}_{\mathrm{RF},u}^{(g)}\right)^H\mathbf{H}^{(g)}_{u, m}\mathbf{F}_{\mathrm{RF},m} \in \mathbb{C}^{N_{\mathrm{RF},u} \times N_{\mathrm{RF},m}}$, and also define the concatenated baseband TPC $\mathbf{f}_{\mathrm{BB}}^{(g)}$ of the $g$th group as $\mathbf{f}_{\mathrm{BB}}^{(g)} = \left[\left(\mathbf{f}_{\mathrm{BB},1}^{(g)}\right)^H, \left(\mathbf{f}_{\mathrm{BB},2}^{(g)}\right)^H, \ldots, \left(\mathbf{f}_{\mathrm{BB},M}^{(g)}\right)^H \right]^H$. The concatenated effective channel matrix between the user $U_{u}^{(g)}$ and all the cooperating APs is denoted as $\mathbf{H}^{(g)}_{\mathrm{eff},u} = [\mathbf{H}^{(g)}_{\mathrm{eff},u,1}, \mathbf{H}^{(g)}_{\mathrm{eff},u,2}, \ldots, \mathbf{H}^{(g)}_{\mathrm{eff},u,m}]\in \mathbb{C}^{N_{\mathrm{RF},u}\times \left(\sum_{m=1}^{M}N_{\mathrm{RF},m}\right)}$. Therefore, the effective received signal after baseband combining, $\hat{\mathbf{r}}_{u}^{(g)}$, at user $U_{u}^{(g)}$ can be written as
\begin{align*}
\hat{\mathbf{r}}_{u}^{(g)} =&\, \left(\mathbf{w}_{\rm BB ,u}^{(g)}\right)^H\mathbf{H}^{(g)}_{\rm eff ,u}\mathbf{f}_{\rm BB}^{(g)} x^{(g)} \nonumber \\ +&\, \left(\mathbf{w}_{\rm BB ,u}^{(g)}\right)^H\sum_{\substack{l=1\\ l \neq m}}^{G}\mathbf{H}^{(g)}_{\rm eff ,k}\mathbf{f}_{\rm BB}^{(l)} x^{(l)} + \left(\mathbf{w}_{\rm BB ,u}^{(g)}\right)^H \tilde{{\boldsymbol{\delta}}}^{(g)}_{u},
\end{align*}
where $\mathbf{w}_{\rm BB ,u}^{(g)} \in \mathbb{C}^{N_{\rm RF, u} \times 1}$ denotes the baseband RC of the user $U_{u}^{(g)}$. The SINR of the user $U_{u}^{(g)}$ can be expressed as

\begin{align*}
\text{SINR}^{(g)}_u =&\, \frac{\left| \left(\mathbf{w}_{\rm BB ,u}^{(g)}\right)^H \mathbf{H}^{(g)}_{\rm eff ,u}\mathbf{f}_{\rm BB}^{(g)}\right|^2}{\sum_{l=1,l \neq m}^{G}{\left| \left(\mathbf{w}_{\rm BB ,u}^{(g)}\right)^H \mathbf{H}^{(g)}_{\rm eff ,u}\mathbf{f}_{\rm BB}^{(l)}\right|^2}+ \sigma_{\tilde{\delta}}^2}.
\end{align*}
Furthermore, the downlink capacity represented as $C_{\mathrm{DL}}$ is given by
\begin{align*}
C_{\mathrm{DL}}=&\, \sum_{g=1}^{G}\sum_{u=1}^{U_g} \log_2 \left(\mbox{1} + \text{SINR}^{(g)}_u\right).
\end{align*}
\subsection{mmWave MIMO channel model}
Using the geometric channel model described in \cite{el2014spatially}, \cite{rappaport2015millimeter} for the mmWave MIMO system, the channel matrix between the $m$th AP and $u$th user, denoted by $\mathbf{H}_{u,m} \in \mathbb{C}^{N_R \times N_T}$, can be expressed as
\begin{equation}
\mathbf{H}_{u,m}=\sqrt{\frac{N_{T}N_{R}}{L}}\sum_{l=1}^{L}\alpha_{u,m}^{(l)}\mathbf{a}_{R}(\phi_{l,u,m}^{R})\mathbf{a}_{T}^H(\phi_{l,u,m}^{T}),
\label{Channel}
\end{equation}
where $L$ represents the number of dominant multipath components between the $m$th AP and user $u$, $\alpha _{u,m}^l$ denotes the complex-valued path gain of the $l\text{th}$ multipath component. Furthermore, the terms $\mathbf{a}_{T}(\phi_{l}^{T})$ and $\mathbf{a}_{R}(\phi_{l}^{R})$ denote the transmit and receive array response vectors, respectively, with $\phi_{l}^{R}$ being the angle of arrival (AoA) and $\phi_{l}^{T}$ the angle of departure (AoD).
For uniformly spaced linear arrays (ULA), where $\mathbf{\lambda}$ and $d$ denote the signal wavelength and distance between the antenna elements, respectively, the array response vectors $\mathbf{a}_{R}(\phi_{l}^{R})$ and $\mathbf{a}_{T}(\phi_{l}^{T})$ are given by
\begin{align*}
& \mathbf{a}_{R}(\phi_{l}^{R}) \\ \nonumber &=\frac{1}{\sqrt{N_{R}}}[1,\ e^{j\frac{2\pi}{\lambda}d\sin(\phi_{l}^{R})},\ \ldots,\ e^{j\frac{2\pi}{\lambda}(N_{R}-1)d\sin(\phi_{l}^{R})}]^{T},
\end{align*}
\begin{align*}
& \mathbf{a}_{T}(\phi_{l}^{T}) \\ \nonumber &=\frac{1}{\sqrt{N_{T}}}[1,\ e^{j\frac{2\pi}{\lambda}d\sin(\phi_{l}^{T})},\ \ldots,\ e^{j\frac{2\pi}{\lambda}(N_{T}-1)d\sin(\phi_{l}^{t})}]^{T}.
\end{align*}
One can also represent this mmWave MIMO channel in a compact form as
\begin{equation*}
\mathbf{H}_{u,m} = \mathbf{A}_{R,u,m}\mathbf{H}_{u,m}^{\alpha}\mathbf{A}_{T,u,m}^{H},
\end{equation*}
where $\mathbf{A}_R = \left[\mathbf{a}_{R}(\phi_{1,u,m}^{R}),\dots,\mathbf{a}_{R}(\phi_{L,u,m}^{R})\right]\in \mathbb{C}^{N_R \times L}$ and $\mathbf{A}_T = \left[\mathbf{a}_{T}(\phi_{1,u,m}^{T}),\dots,\mathbf{a}_{T}(\phi_{L,u,m}^{T})\right]\in \mathbb{C}^{N_T \times L}$ contain the array response vectors of the AP and user, respectively, and $\mathbf{H}_{u,m}^{\alpha} = \mathrm{diag}\left(\alpha_{u,m}^{(1)},\alpha_{u,m}^{(2)},\dots,\alpha_{u,m}^{(L)}\right)$ denotes the diagonal matrix of path gains. Although this channel model is described for a uniform linear array (ULA), one can also readily extend the framework of this paper to uniform planar arrays (UPA). The algorithms and results derived in this work are also valid for a UPA, the corresponding array response vector can be written as
\begin{align}
 \mathbf{a}_{x}\left(\phi,\theta\right)&= \frac{1}{\sqrt{N_x}}\bigg[1, e^{j\frac{2\pi}{\lambda}d\left(\sin\phi \sin\theta + \cos\theta \right)}, \ldots, \nonumber \\ & \ldots, e^{j\frac{2\pi}{\lambda}d\left(\left(N_{x}^{h}-1\right)\sin\phi \sin\theta + \left(N_{x}^{v}-1\right) \cos\theta \right)} \bigg],
\end{align}
where $x \in \lbrace R,T \rbrace$, while $N_{x}^{h}$ and $N_{x}^{v}$ denote the number of horizontal and vertical elements of the UPA in the two-dimension plane of the array. The UPA structure at each AP has a total of $N_T$ elements associated with $N_T = N_{T}^{h} \times N_{T}^{v}$.

Note that the system model developed in Section-II describes a general multicast scenario, where a group of users may request identical services. A special case of the multicast scenario is popularly termed as a unicast scenario in which each user requests different information. The next section designs HTBFs for a broadcast system, where all the users request the same information, i.e., there is only a single group. Thus, it can also be viewed as a special instance of the multicast scenario, where there is no inter-group or inter-user interference. This interference-free scenario simplifies the HTBF design problem, as described in the next section.
\section{Hybrid Broadcast Beamforming (HBBF)}\label{HBB}
\label{sec HBB}
In a broadcast scenario, the sum signal-to-noise ratio (SSNR) of all the users in the system is maximized subject to the total power constraint ($P_T$) in order to obtain the optimal HBBF. The motivation for SSNR maximization can be summarised as follows. Note that at low-SNRs, we have $\log_2 \left(1 + \mathrm{SNR}_u\right) \approx \frac{1}{\mathrm{ln}\left(2\right)}\mathrm{SNR}_u$. Therefore, one can deduce that the maximization of capacity is identical to SSNR maximization. Furthermore, it can observed from the following relationship, which exploits Jensen's inequality, that SSNR maximization provides an upper bound of the true capacity:
$$\frac{1}{U} \sum_{u=1}^U \log_2 \left(1 + \mathrm{SNR}_u\right) \leq \log_2 \left(1 + \frac{1}{U}\sum_{u=1}^U\right).$$
Considering a broadcast scenario to $U$ users, the signal $\mathbf{r}_u \in \mathbb{C}^{N_R \times 1} $ received at user $u$ can be expressed as
\begin{equation*}
\mathbf{r}_u = \sum_{m=1}^{M}\mathbf{H}_{u,m}\mathbf{F}_{\mathrm{RF},m}\mathbf{f}_{\mathrm{BB},m}x + \boldsymbol{\delta}_u,
%\label{Equation 1}
\end{equation*}
where $\mathbf{H}_{u,m}\in \mathbb{C}^{N_{R} \times N_{T}} $ represents the channel matrix between the user $u$ and the $m$th AP, $\mathbf{F}_{\mathrm{RF},m} \in \mathbb{C}^{N_{T} \times N_{\mathrm{RF},m}}$ and $\mathbf{f}_\mathrm{BB,m} \in \mathbb{C}^{N_{\mathrm{RF},m} \times 1}$ denote the RF and baseband TPCs respectively, at the $m$th AP. Furthermore,  $\boldsymbol{\delta}_u \in \mathbb{C}^{N_{R} \times 1}$ represents the AWGN at user $u$ and $x$ denotes the transmitted symbol corresponding to the broadcast scenario, which is normalized to have unit power i.e., $ \mathop{\mathbb{E}}\left[|x|^2 \right] = 1$. After RF combining using the combiner matrix $\mathbf{W}_{\mathrm{RF},u} \in \mathbb{C}^{N_R \times N_{\mathrm{RF},u}}$, 
the received signal $\tilde{\mathbf{r}}_u$ can be written as
\begin{align}
\tilde{\mathbf{r}}_u =& \mathbf{W}_{\mathrm{RF},u}^H \left( \sum_{m=1}^{M}\mathbf{H}_{u,m}\mathbf{F}_{\mathrm{RF},m} \mathbf{f}_{\mathrm{BB},m}x \right) + \tilde{\boldsymbol{\delta}}_u,
\label{Equation 2}
\end{align}
where $\tilde{\boldsymbol{\delta}}_u = \mathbf{W}_{\mathrm{RF},u}^H \boldsymbol{\delta}_u$. Note that the columns of the RF TPC $\mathbf{F}_{\mathrm{RF},m}$ can be chosen as the dominant $N_{\mathrm{RF},m}$ transmit array steering vectors corresponding to each user. Similarly, the RF RC $\mathbf{W}_{\mathrm{RF},u}$ can be configured with the $N_{\mathrm{RF},u}$ dominant receive array steering vectors corresponding to each AP. Thus, to design the RF precoder for the $m$th AP, one can first find the index $\mathcal{K}_{u,m}$ of the multipath component corresponding to the $u$th user, which has the maximum magnitude path gain, i.e., $\mathcal{K}_{u,m} = \max\left\{\left|\alpha_{u,m}^{(l)}\right|, \left|\alpha_{u,m}^{(2)}\right|,\dots,\left|\alpha_{u,m}^{(L)}\right|\right\}$. Then, the RF TPC $\mathbf{F}_{\mathrm{RF},m}$ for the $m$th AP
%and the RF combiner for the $u$th user $\mathbf{W}_{\mathrm{RF},u}$
can be obtained by extracting the $N_{\mathrm{RF},m}$ columns indexed by the set $\mathcal{K}_m = \left\{\mathcal{K}_{1,m},\mathcal{K}_{2,m},\dots,\mathcal{K}_{U,m}\right\}$ from the matrices $\mathbf{A}_{T,u,m}$. This can be mathematically represented as
\begin{equation}
\mathbf{F}_{\mathrm{RF},m} = \left[\mathbf{A}_{T,1,m}(:,\mathcal{K}_{1,m}),\dots, \mathbf{A}_{T,U,m}(:,\mathcal{K}_{U,m})\right].
\end{equation}  

In a similar fashion, one can also obtain the RF RC $\mathbf{W}_{\mathrm{RF},u}$ for each user. Let the effective channel matrix $\mathbf{H}_{\mathrm{eff},u} \in \mathbb{C}^{N_{\mathrm{RF},u} \times N_{\mathrm{RF},m}}$ between the user $u$ and the $m$th AP be defined as
$\mathbf{H}_{\mathrm{eff},u,m} = \mathbf{W}_{\mathrm{RF},u}^H\mathbf{H}_{u,m}\mathbf{F}_{\mathrm{RF},m}.$
Furthermore, let the concatenated effective channel matrix $\mathbf{H}_{\mathrm{eff},u} \in \mathbb{C}^{N_{\mathrm{RF},u} \times \sum_{m=1}^{M}N_{\mathrm{RF},m}}$ for user $u$ be defined as
$$\mathbf{H}_{\mathrm{eff},u} = \left[\mathbf{H}_{\mathrm{eff},u,1} \mathbf{H}_{\mathrm{eff},u,2} \cdots \mathbf{H}_{\mathrm{eff},u,M}  \right].$$ 
The received signal $\tilde{\mathbf{r}}_u$ of \eqref{Equation 2} can be completely rewritten as
\begin{align}\label{EqBHBRBroad}
&\tilde{\mathbf{r}}_u =  \sum_{m=1}^{M}\mathbf{H}_{\mathrm{eff},u,m}\mathbf{f}_{\mathrm{BB},m}x + \tilde{\boldsymbol{\delta}}_ u = \mathbf{H}_{\mathrm{eff},u} \mathbf{f}_\mathrm{{BB}}x + \tilde{\boldsymbol{\delta}}_ u,
\end{align}
where $\mathbf{f}_\mathrm{{BB}} = \left[\mathbf{f}_{\mathrm{BB},1}^T, \mathbf{f}_{\mathrm{BB},2}^T, \cdots, \mathbf{f}_{\mathrm{BB},M}^T\right]^T \in \mathbb{C}^{\sum_{m=1}^{M} N_{\mathrm{RF},m} \times 1}$ denotes the stacked baseband TPs of all the APs. Note that the  signal power received at user $u$ is given as $\mathbf{f}_\mathrm{BB}^H\mathbf{H}_{\mathrm{eff},u}^H\mathbf{H}_{\mathrm{eff},u}\mathbf{f}_\mathrm{BB} $, which implies that the total received power for all the users in the system can be written as $\sum_{u=1}^{U}\mathbf{f}_\mathrm{BB}^H\mathbf{H}_{\mathrm{eff},u}^H\mathbf{H}_{\mathrm{eff},u}\mathbf{f}_\mathrm{BB} = \mathbf{f}_\mathrm{BB}^H  \mathbf{H}\mathbf{f}_\mathrm{BB}$, 
where $\mathbf{H} = \sum_{u=1}^{U}\mathbf{H}_{\mathrm{eff},u}^H\mathbf{H}_{\mathrm{eff},u} $. Our objective is to find the optimal stacked baseband TPC $\mathbf{f}_\mathrm{BB}$ for the HBBF in order to maximize the SSNR. Toward this, the optimization problem can be formulated as
\begin{equation*}
\begin{aligned}
& \underset{\mathbf{f}_\mathrm{BB}}{\text{maximize}}
& & \mathbf{f}_\mathrm{BB}^H  \mathbf{H}\mathbf{f}_\mathrm{BB} \\
& \text{subject to}
& & \left\lVert \mathbf{f}_\mathrm{BB} \right\rVert^2 \leq P_T.
\end{aligned}
\end{equation*} 

The optimal baseband TPC $\mathbf{f}_\mathrm{BB}$ for the optimization problem above is given by $\mathbf{f}_\mathrm{BB}=  \sqrt{P_T}  \boldsymbol{\lambda}_{ \text{max}} \left(\mathbf{H}\right),$
where $\boldsymbol{\lambda}_{ \text{max}}(\mathbf{H})$ represents the eigenvector corresponding to the largest eigenvalue of the matrix $\mathbf{H}$. Interestingly, it can be observed that the SSNR of the mmWave broadcast system can be maximized by sending the information in the direction of the leading eigenmode of the effective channel matrix $\mathbf{H}$.
Substituting this optimal $\mathbf{f}_{\mathrm{BB}}$ into \eqref{EqBHBRBroad}, we have
\begin{equation}
\tilde{\mathbf{r}}_u = \sqrt{P_T} \mathbf{H}_{\mathrm{eff},u}   \boldsymbol{\lambda}_{ \text{max}} \left(\mathbf{H}\right)x + \tilde{\boldsymbol{\delta}}_u .
\end{equation}
The covariance matrix $\mathbf{R}_{\tilde{\delta}_ u} \in \mathbb{C}^{N_{\mathrm{RF},u} \times N_{\mathrm{RF},u}}$ of the combined noise vector $\tilde{\boldsymbol{\delta}}_u $ can be expressed as
\begin{equation}
 \mathbf{R}_{\tilde{\delta}_ u}= \mathop{\mathbb{E}}\left[\tilde{\boldsymbol{\delta}}_ u\tilde{\boldsymbol{\delta}}_ u^{H}\right] = \sigma^2_{\delta} \mathbf{W}_{\mathrm{RF},u}^H\mathbf{W}_{\mathrm{RF},u}.
\end{equation}
Therefore, the whitened system model can be written as
\begin{equation}
\mathbf{R}_{\tilde{\delta}_ u}^{-\frac{1}{2}} \tilde{\mathbf{r}}_u = \sqrt{P_T} \mathbf{R}_{\tilde{\delta}_ u}^{-\frac{1}{2}} \mathbf{H}_{\mathrm{eff},u}   \boldsymbol{\lambda}_{ \text{max}} \left(\mathbf{H}\right)x + \mathbf{R}_{\tilde{\mathbf{\delta}}_ u}^{-\frac{1}{2}}\tilde{\boldsymbol{\delta}}_ u .
\end{equation}
The optimal baseband RC $\mathbf{w}_{\mathrm{BB},u}$  for user $u$ can now be readily found using the MRC criterion as
\begin{equation}
\mathbf{w}_{\mathrm{BB},u} = \frac{\mathbf{R}_{\tilde{\delta}_ u}^{-\frac{1}{2}}\mathbf{H}_{\mathrm{eff},u}   \boldsymbol{\lambda}_{ \text{max}} \left(\mathbf{H}\right)}{\left\lVert \mathbf{R}_{\tilde{\delta}_ u}^{-\frac{1}{2}}\mathbf{H}_{\mathrm{eff},u}   \boldsymbol{\lambda}_{ \text{max}} \left(\mathbf{H}\right)  \right\rVert}.
\end{equation}
Next, we derive the cooperative HBBF for a broadcast system having per AP-power constraints. Furthermore, we also maximize the minimum SNR of the users in the system to incorporate fairness in the power allocation.
\subsection{HBBF-P}
This sub-section considers per-AP power constraints for performing cooperative HBBF for CFMU systems that maximizes the sum-SNR. Let us assume the power among the APs to be equally distributed as $\frac{P_T}{M}$ in the broadcast group. The associated optimization problem of the CFMU HBBF, which maximizes the sum-SNR, is given by \vspace{-10pt}

\begin{align*}
& \underset{\mathbf{F}_{\mathrm{BB},u}}{\text{maximize}}\quad \sum_{u=1}^{U} \text{Tr} \left(\mathbf{f}_\mathrm{BB}^H \mathbf{H}_{\mathrm{eff},u}^H\mathbf{H}_{\mathrm{eff},u}\mathbf{f}_\mathrm{BB}\right)\\ &\qquad\quad\ \equiv \sum_{u=1}^{U}\sum_{m=1}^{M} \text{Tr} \left(\mathbf{H}_{\mathrm{eff},u,m}^H\mathbf{H}_{\mathrm{eff},u,m}\mathbf{F}_{\mathrm{BB},m}\right)\\
& \text{subject to} \ \text{Tr}\left(\mathbf{F}_{\mathrm{BB},m}\right) \leq \frac{P_{T}}{M}, \, m=\mbox{1}, \mbox{2}, \ldots, M \nonumber \\ &\qquad\quad\ \ \ \ \ \mathbf{F}_{\mathrm{BB},m}\succcurlyeq 0 \ \ \text{and} \ \text{rank}\left(\mathbf{F}_{\mathrm{BB},m}\right)=1 \ \ \ \forall m,
\end{align*}
where $\mathbf{F}_{\mathrm{BB},m}=\mathbf{f}_{\mathrm{BB},m}\left(\mathbf{f}_{\mathrm{BB},m}\right)^H \in \mathbb{C}^{N_{\mathrm{RF},m} \times N_{\mathrm{RF},m}} $ denotes a rank-1 positive semi-definite (PSD) matrix. The above optimization problem is non-convex due to the rank-1 constraint, hence it is difficult to solve. Upon relaxing this rank-1 constraint, the equivalent semi-definite programme (SDP) can be expressed as
\begin{equation}\label{HBBP}
\begin{aligned}
& \underset{\mathbf{F}_{\mathrm{BB},m}}{\text{maximize}}
& &  \sum_{u=1}^{U}\sum_{m=1}^{M} \text{Tr} \left(\mathbf{H}_{\mathrm{eff},u,m}^H\mathbf{H}_{\mathrm{eff},u,m}\mathbf{F}_{\mathrm{BB},m}\right)\\
& \text{subject to}
& & \text{Tr}\left(\mathbf{F}_{\mathrm{BB},m}\right) \leq \frac{P_{T}}{M}, \ \ \    m=1,2, \cdots, M, \\
& & & \ \mathbf{F}_{\mathrm{BB},m}\succcurlyeq 0 \ \ \ \forall m.
\end{aligned}
\end{equation}
\subsection{HBBF-P with a max-min per-AP power constraint}

In order to ensure fairness in the power allocation, the minimum SNR of the users can be maximized. Towards this, let the target SNR of each user be denoted by $\gamma$. To perform HBBF, the max-min fairness based optimization problem can be framed as
\begin{equation} \label{HBB max-min}
\begin{aligned}
\ \ \ \underset{\gamma, \mathbf{F}_\mathrm{BB,n}}{\text{maximize}}
\ \gamma \\
\hspace{-3cm} \text{s.t.}\ \ 
& \sum_{m=1}^{M} \text{Tr} \left(\mathbf{H}_{\mathrm{eff},u,m}^H\mathbf{H}_{\mathrm{eff},u,m}\mathbf{F}_{\mathrm{BB},m}\right) \geq \gamma, \ \forall u, \\
& \text{Tr}\left(\mathbf{F}_{\mathrm{BB},m}\right) \leq \mathrm{\frac{P_{T}}{M}}, \ \ \ \    m=1,2, \cdots, M, \\
& \mathbf{F}_{\mathrm{BB},m}\succcurlyeq 0, \ \ \ \forall m.
\end{aligned}
\end{equation}
It can be observed that the optimization problems \eqref{HBBP} and \eqref{HBB max-min}  can be efficiently solved using widely available tools, such as CVX \cite{grant2014cvx}. The baseband beamformer $\mathbf{f}_{\mathrm{BB},m}$ can be finally determined as the eigenvector corresponding to the largest eigenvalue of $\mathbf{F}_{\mathrm{BB},m}^{*}$, i.e., $\mathbf{f}_{\mathrm{BB},m}^{\mathrm{opt}} = \sqrt{\lambda}_{\mathrm{max}}\hat{\mathbf{f}}_{\mathrm{BB},m}$, where $\hat{\mathbf{f}}_{\mathrm{BB},m}$ represents the unit-norm eigenvector corresponding to the eigenvalue $\lambda_{\mathrm{max}}$. Note that the relaxed SDP based beamformer obtained is an approximation of the optimal beamformer. It has been shown that the SDP relaxation results in a close approximation of the optimization problems involving non-convex quadratically constrained quadratic programs (QCQPs) \eqref{HBBP} and max–min-based quadratic programs \eqref{HBB max-min}.
Furthermore, it has been demonstrated that for SDP relaxation, the rank of an optimal solution is upper bounded by $\sqrt{M}$ \cite{palomar2010convex}. For practical values of $M$, the experimental studies in \cite{palomar2010convex} demonstrate that the average approximation ratio is close to unity. The computational complexity order of HBBF-P scheme is $\mathcal{O}\left[\left(MN_{\mathrm{RF},m}\right)^{3.5}\right]$ \cite{luo2010semidefinite, vandenberghe1996semidefinite}.
\section{Unicast Scenario: Optimal Successive Hybrid Transmit Beamforming}\label{Unicast}
This section jointly designs the optimal HTBFs and RCs for the mmWave unicast scenarios by maximizing the SINR in successive order. The algorithm developed herein will be extended to multicast systems in the next section.
In this setting, there are $G = U$ user groups with one user in each group, i.e., $U_g = 1$, where $U$ denotes the total number of users present in the system. The signal $\mathbf{r}_u \in \mathbb{C}^{N_R \times 1}$ received at the $u$th user can be written as
\begin{align}
 \mathbf{r}_u =& \sum_{m=1}^{M}\mathbf{H}_{u,m}\mathbf{F}_{\mathrm{RF},m}\mathbf{f}_{\mathrm{BB},u,m}x_u \nonumber \\ +& \sum_{\substack{i=1\\ i \neq u}}^{U} \sum_{m=1}^{M}\mathbf{H}_{u,m}\mathbf{F}_{\mathrm{RF},m}\mathbf{f}_{\mathrm{BB},i,m}x_i + \boldsymbol{\delta}_{u},
\label{Equation u1}
\end{align}
where $\boldsymbol{\delta}_u \in \mathbb{C}^{N_{R} \times 1}$ represents the AWGN at user $u$ and $x_u$ denotes the symbol transmitted to user $u$, which is normalized to have unit power. Furthermore, $\mathbf{f}_{\mathrm{BB},u,m} \in \mathbb{C}^{N_{\mathrm{RF},m} \times 1}$ represents the baseband TPC at the $m$th AP for the user $u$.
After employing the RF RC $\mathbf{W}_{\mathrm{RF},u} \in \mathbb{C}^{N_{u} \times N_{\mathrm{RF},u}}$, the received signal, $\tilde{\mathbf{r}}_u = \mathbf{W}_{\mathrm{RF},u}^H \mathbf{r}_u$, is given as:
\begin{align}\label{EqBHBR}
\tilde{\mathbf{r}}_u =&\, \mathbf{W}_{\mathrm{RF},u}^H\sum_{m=1}^{M}\mathbf{H}_{u,m}\mathbf{F}_{\mathrm{RF},m}\mathbf{f}_{\mathrm{BB},u,m}x_u \nonumber \\ +& \mathbf{W}_{\mathrm{RF},u}^H\sum_{\substack{i=1\\ i \neq u}}^{U} \sum_{m=1}^{M}\mathbf{H}_{u,m}\mathbf{F}_{\mathrm{RF},m}\mathbf{f}_{\mathrm{BB},i,m}x_i + \tilde{\boldsymbol{\delta}}_u,
\end{align}
where $\tilde{\boldsymbol{\delta}}_u = \mathbf{W}_{\mathrm{RF},u}^H \boldsymbol{\delta}_u$. Let $\mathbf{H}_{\mathrm{eff},u,m} = \mathbf{W}_{\mathrm{RF},u}^H\mathbf{H}_{u,m}\mathbf{F}_{\mathrm{RF},m}$ denote the effective channel matrix between the $m$th AP and user $u$. 
Let the concatenated baseband TPC and the effective channel matrix of the $u$th user be defined as
$$\mathbf{f}_{\mathrm{BB},u} = [\mathbf{f}_{\mathrm{BB},u,1}^{T}, \mathbf{f}_{\mathrm{BB},u,2}^{T}, \ldots, \mathbf{f}_{\mathrm{BB},u,M}^{T}]^{T},$$ 
$$\!\!\!\!\!\!\!\!\! \mathbf{H}_{\mathrm{eff},u} = \left[\mathbf{H}_{\mathrm{eff},u,1} \mathbf{H}_{\mathrm{eff},u,2} \cdots \mathbf{H}_{\mathrm{eff},u,M}  \right],$$ respectively.
After baseband combining at user $u$ employing the RC $\mathbf{w}_{\mathrm{BB},u} \in \mathbb{C}^{N_{\mathrm{RF}, u} \times 1}$, the estimate $\hat{x}_u$ of the symbol $x_u$ is given by
\begin{align}\label{estUnicast}
\hat{x}_u =& \mathbf{w}_{\mathrm{BB},u}^H\mathbf{H}_{\mathrm{eff},u} \mathbf{f}_{\mathrm{BB},u} x_u + \sum_{i=1, i \neq u}^{U}\mathbf{w}_{\mathrm{BB},u}^H \mathbf{H}_{\mathrm{eff},u} \mathbf{f}_{\mathrm{BB},i} x_i \nonumber \\ +&  \mathbf{w}_{\mathrm{BB},u}^H \boldsymbol{\delta}_{u}.
\end{align}
Note that the first and second terms in \eqref{estUnicast} correspond to the desired signal and inter-user interference, respectively. Next, we develop the OSHB-U scheme similar to the successive optimization method in multi-user MIMO systems \cite{spencer2004zero} that employs cooperative beamforming for iteratively multiplexing multiple users in the downlink, together with exploiting the MVDR beamformer of \cite{van2002modulation} for interference minimization in the mmWave cellular system. Furthermore, it maximizes the SINR of each user while cancelling the interferences imposed on the users, which have already been scheduled. Toward this, we now derive the proposed OSHB-U technique for a unicast scenario.

To begin with, it can be readily observed that while scheduling the $u$th user, the zero-forcing (ZF) interference constraint of $\mathbf{w}_{\mathrm{BB},v}^H \mathbf{H}_{\mathrm{eff},v} \mathbf{f}_{\mathrm{BB},u} = \mbox{0}$, $\mbox{1}\leq v \leq u-\mbox{1}$, successfully eliminates the interference imposed by the previously scheduled users. Hence, the baseband TPC is given by $\mathbf{f}_{\mathrm{BB},u}=\mathbf{M}_{u-1}^{\perp} \boldsymbol{\nu}_{u}$, where $\boldsymbol{\nu}_{u} \in \mathbb{C}^{\sum_{m=1}^{M}N_{\mathrm{RF},m}-u+1}$, and $\mathbf{M}_{u-1}^{\perp}\in\mathbb{C}^{\sum_{m=1}^{M}N_{\mathrm{RF},m}\times(\sum_{m=1}^{M}N_{\mathrm{RF},m}-u+1)}$ represents the orthonormal basis to the null space of the matrix $\mathbf{M}_{u-1} \in \mathbb{C}^{(u-1) \times \sum_{m=1}^{M} N_{\mathrm{RF},m}}$, which is given by
\begin{align}\label{MainMatrix2}
\mathbf{M}_{u-1}=\left[\left(\mathbf{w}_{\mathrm{BB},1}^H\mathbf{H}_{\mathrm{eff},1}\right)^T,\ldots, \left(\mathbf{w}_{\mathrm{BB},u-1}^H\mathbf{H}_{\mathrm{eff},u-1}\right)^T\right]^T.
\end{align}
The matrix $\mathbf{M}_{u-1}^{\perp}$ can be readily derived from the singular value decomposition (SVD) of the matrix $\mathbf{M}_{u-1}$, and it corresponds to the right singular vectors matrix composed of $\left(\sum_{m=1}^{M}N_{\mathrm{RF},m}-u+1\right)$. Furthermore, the transmissions of the successive users $u+1, u+2, \ldots, U$ are also orthogonal to those of the $u$th user, i.e., $\mathbf{w}_{\mathrm{BB},u}^H \mathbf{H}_{\mathrm{eff},u} \mathbf{f}_{\mathrm{BB},v}=\mbox{0}$, $u + 1 \leq v \leq U$. Therefore , the effective received signal after RF precoding and combining at user $u$ can be expressed as
\begin{align}
\tilde{\mathbf{r}}_u =&\, \mathbf{H}_{\mathrm{eff}, u} \mathbf{f}_{\mathrm{BB},u} x_u + \sum_{i=1}^{u-1}\mathbf{H}_{\mathrm{eff}, u} \mathbf{f}_{\mathrm{BB},i} x_i + \tilde{\boldsymbol{\delta}}_{u} \nonumber \\ =&\, \mathbf{H}_{\mathrm{eff},u} \mathbf{M}_{u-1}^{\perp} \boldsymbol{\nu}_{u} x_u + \sum_{i=1}^{u-1}\mathbf{H}_{\mathrm{eff},u} \mathbf{M}_{i-1}^{\perp} \boldsymbol{\nu}_{i} x_i + \tilde{\boldsymbol{\delta}}_{u}.
\label{eq 16}
\end{align}
Let the interference and noise terms in \eqref{eq 16} be collectively denoted by $\boldsymbol{\tau}_u = \sum_{i=1}^{u-1}\mathbf{H}_{\mathrm{eff},u} \mathbf{M}_{u-1}^{\perp} \boldsymbol{\nu}_{i} x_i + \tilde{\boldsymbol{\delta}}_{u} $. Its covariance matrix $\mathbf{C}_{\boldsymbol{\tau}_{u}} \in \mathbb{C}^{N_{\mathrm{RF},u} \times N_{\mathrm{RF},u}}$ is given by

\begin{align}\label{covar matrix}
\mathbf{C}_{\boldsymbol{\tau}_{u}} =&\, \sum_{i=1}^{u-1}\mathbf{H}_{\mathrm{eff},u} \mathbf{f}_{\mathrm{BB},i} \left(\mathop{\mathbb{E}}\left[|x_i|^2 \right]\right)\mathbf{f}_{\mathrm{BB},i}^{H} \mathbf{H}_{\mathrm{eff},u}^{H} + \mathop{\mathbb{E}}\left[\boldsymbol{\delta}_{u} \boldsymbol{\delta}_{u}^H \right]\nonumber \\ =&\, \sum_{i=1}^{u-1} \mathbf{H}_{\mathrm{eff},u} \left(\mathbf{M}_{i-1}^{\perp} \boldsymbol{\nu}_{i} \boldsymbol{\nu}_{i}^{H} \left(\mathbf{M}_{i-1}^{\perp}\right)^{H}\right) \mathbf{H}_{\mathrm{eff},u}^{H} \nonumber \\ +& \sigma_{\boldsymbol{\delta}}^{2}\mathbf{W}_{\mathrm{RF},u}^H\mathbf{W}_{\mathrm{RF},u}.
\end{align}
To obtain the optimal baseband TPC and RC, one can write the effective interference-whitened system model corresponding to (\ref{eq 16}) as
\begin{equation}
\mathbf{C}_{\boldsymbol{\tau}_{u}}^{-\frac{1}{2}}\tilde{\mathbf{r}}_{u}= \mathbf{C}_{\boldsymbol{\tau}_{u}}^{-\frac{1}{2}}\mathbf{H}_{\mathrm{eff},u}\mathbf{M}_{u-1}^{\perp} \boldsymbol{\nu}_{u} x_{u} +\mathbf{C}_{\boldsymbol{\tau}_{u}}^{-\frac{1}{2}}\boldsymbol{\tau}_{u}.
\end{equation}
Therefore, the optimal baseband TPC and RC for user $u$, which maximize the SINR are given by $\mathbf{f}_{\mathrm{BB},u} = \mathbf{M}_{u-1}^{\perp} \boldsymbol{\nu}_{u}$  and $\mathbf{w}_{\mathrm{BB},u} = {\left(\mathbf{C}_{\boldsymbol{\tau}_{u}}^{-(1/2)}\right)}^H \tilde{\mathbf{w}}_{\mathrm{BB},u}$, where $\tilde{\mathbf{w}}_{\mathrm{BB},u}$ and $\boldsymbol{\nu}_{u}$ are the principal left and right singular vectors, respectively, of the matrix $\mathbf{C}_{{\boldsymbol{\tau}}_{u}}^{-({1}/{2})}\mathbf{H}_{\mathrm{eff},u}\mathbf{M}_{u-1}^{\perp} \in \mathbb{C}^{N_{\mathrm{RF},u} \times (MN_{\mathrm{RF},m} - u+1)}$. The detailed proofs of the results obtained above are given in Appendix A.
The optimal vector $\mathbf{w}_{\mathrm{BB},u} = {\left(\mathbf{C}_{\boldsymbol{\tau}_{u}}^{-(1/2)}\right)}^H \tilde{\mathbf{w}}_{\mathrm{BB},u}$ is known as the MVDR beamformer that maximizes the SINR. This is termed as MVDR, because the output signal at user $u$ has the minimum variance and the desired signal is not distorted.
Therefore, the minimum variance distortionless response (MVDR) together with the optimal successive minimum variance hybrid beamforming (OSHB) results in the optimal TPC that maximizes the SINR, while also nulling the interference emanating from the previously scheduled $u-1$ users. Note that the proposed optimal hybrid beamformer is obtained in two steps. In the first step, we employ the CSI toward design of the optimal RF TPCs. In the second step, the proposed baseband TPC is derived based on the MVDR beamformer. Therefore, this results in an optimal hybrid beamformer that maximizes the SINR along with interference cancellation.
Algorithm \ref{algo_S_OMP} presents the step-by-step procedure of the OSHB algorithm for obtaining the baseband TPC and RC. The key features of the proposed OSHB technique for the mmWave unicast CFMU MIMO system are described next.

To begin with, note that the interference cancellation procedure of the proposed OSHB method is similar to the classic successive interference cancellation (SIC) methods, wherein once the signal of user $u$ is successfully recovered, one can remodulate it and then subtract it from the received composite signal for decontaminating it and for reducing the decoding complexity of the remaining users at the APs. Hence, the baseband TPC configured for the $u$th user is specifically chosen for ensuring that the interference emanating from the formerly selected $(u-1)$ users is zero. This can be realized by selecting the TPC for ensuring that it lies in the subspace orthogonal to the column space of the matrix $\mathbf{M}_{u-1}$.
Furthermore, one can observe that the maximum number of users supported by the OSHB-U scheme is $\sum_{m=1}^{M}N_{\mathrm{RF},m}$. In order to cancel the interference imposed on user $u$, the number of columns in the null space of the matrix $\mathbf{M}_{u-1}^{\perp}\in\mathbb{C}^{\sum_{m=1}^{M}N_{\mathrm{RF},m}\times(\sum_{m=1}^{M}N_{\mathrm{RF},m}-u+1)}$ must be larger than or equal to one, i.e. $\sum_{m=1}^{M}N_{\mathrm{RF},m}-u+1 \geq 1$, which yields $U \leq \sum_{m=1}^{M}N_{\mathrm{RF},m}$. This implies that the number of RF chains at each AP has to be increased for multiplexing a higher number of users. On the other hand, if one considers the overall effective channel matrix $\mathbf{H}_\mathrm{eff} = {[\mathbf{H}_{\mathrm{eff},1}^T,\mathbf{H}_{\mathrm{eff},u}^T, \ldots,\mathbf{H}_{\mathrm{eff},U}^T]}^T \in \mathbb{C}^{UN_{\mathrm{RF},u} \times M N_{\mathrm{RF},m}}$ to perform block diagonalization, then one has to satisfy the constraint $UN_{\mathrm{RF},u} \leq \sum_{m=1}^{M}N_{\mathrm{RF},m}$. Hence, the block diagonalization method specifically designed for perfect interference removal can only support $\lfloor \sum_{m=1}^{M} N_{\mathrm{RF},m}/N_{\mathrm{RF},u}\rfloor$ users. Therefore, it can be readily observed that a substantially higher number of users can be served using the proposed OSHB scheme than by the conventional block diagonalization approach, which demonstrates its practical suitability.

\begin{algorithm}[t!]
\caption{OSHB algorithm}\label{algo_S_OMP}
\textbf{Input:} Effective channel matrix  $\mathbf{H}_{\mathrm{eff},u}$, $\mathbf{W}_{\mathrm{RF},u}$\\
\textbf{Initialization:} $\mathbf{M}_{0} =  \mathbf{I}_{1 \times MN_{\mathrm{RF},m}}$, $\mathbf{C}_{\tau_{1}} = \mathbf{W}_{\mathrm{RF},1}^H\mathbf{W}_{\mathrm{RF},1}$ \\
\textbf{for} $u = 1:U$ \textbf{do}
\begin{enumerate}
\item $\left[\mathbf{U} \ \mathbf{S} \ \mathbf{V}\right] = \mathrm{svd}\left(\mathbf{M}_{u-1}\right)$
\item $\mathbf{M}_{u-1}^{\perp} = \mathbf{V}\left(:,u:MN_{\mathrm{RF},m}\right)$
\item $\left[\mathbf{U}^{'} \ \mathbf{S}^{'} \ \mathbf{V}^{'} \right] = \mathrm{svd}\left(\mathbf{C}_{\tau_{u}}^{\frac{-1}{2}}\mathbf{H}_{\mathrm{eff},u}\mathbf{M}_{u-1}^{\perp}\right)$
\item $\boldsymbol{\nu}_{u} = \mathbf{V}^{'}\left(:,1\right)$

\item $\mathbf{f}_{\mathrm{BB},u} = \mathbf{M}_{u-1}^{\perp}\boldsymbol{\nu}_{u}$
\item $\mathbf{w}_{\mathrm{BB},u} = \left(\mathbf{C}_{\tau_{u}}^{\frac{-1}{2}}\right)^H\mathbf{U}^{'}\left(:,1\right)$
\item $\mathbf{C}_{\tau_{u}} \rightarrow $ update according to \eqref{covar matrix}
\item $\mathbf{M}_{u-1} \rightarrow $ update according to \eqref{MainMatrix2}   
\end{enumerate}
\textbf{end for}\\
\end{algorithm}

\section{Optimal Successive Hybrid Beamforming for Multicast Scenarios}
\label{Multicastsytem}
As described previously, in a multicast scenario, the users belonging to the same multicast group tend to request the same data. For such a system, one can develop the OSHB-M scheme, similar to the OSHB-U scheme derived for the unicast scenario, which successfully cancels the interference arising from the previously chosen groups $1, 2,\ldots, g-1$ contaminating the $g$th group. Furthermore, to minimize the interference generated by the previously chosen groups, the MVDR-based beamformer can be employed for the $g$th group.

As described in Section II, the signal $\tilde{\mathbf{r}}^{(g)}_u \in \mathbb{C}^{N_{\mathrm{RF},u} \times 1}$ received for the user $U_u^{(g)}$ at the output of the RF RC is given as
\begin{align}
\tilde{\mathbf{r}}^{(g)}_u = \mathbf{H}^{(g)}_{\mathrm{eff},u}\mathbf{f}_{\mathrm{BB}}^{(g)} x^{(g)} + \sum_{\substack{l=1\\ l \neq i}}^{G}\mathbf{H}^{(g)}_{\mathrm{eff},u}\mathbf{f}_{\mathrm{BB}}^{(l)} x^{(l)} + \check{{\boldsymbol{\delta}}}^{(g)}_{u}.
\end{align}
The received signal concatenated for all the users in group $g$ can be formulated as
\begin{align}
\underset{{\mathbf{r}^{(g)}}}{\underbrace{\begin{bmatrix}\mathbf{r}_{1}^{(g)} \\ \mathbf{r}_{2}^{(g)} \\ \vdots \\ \mathbf{r}_{U_{g}}^{(g)} \end{bmatrix}}}=\underset{\mathbf{H}^{(g)}_{\mathrm{eff}}}{\underbrace{\begin{bmatrix} \mathbf{H}^{(g)}_{\mathrm{eff},1} \\ \mathbf{H}^{(g)}_{\mathrm{eff},2} \\ \vdots \\ \mathbf{H}^{(g)}_{\mathrm{eff},U_{g}} \end{bmatrix}}}\mathbf{f}_{\mathrm{BB}}^{(g)} x^{(g)}+\sum_{\substack{l=1\\ l \neq g}}^{G} \mathbf{H}^{(g)}_{\mathrm{eff}}\mathbf{f}_{\mathrm{BB}}^{(l)} x^{(l)}+ \underset{\boldsymbol{\delta}^{(g)}}{\underbrace{\begin{bmatrix} \boldsymbol{\delta}_{1}^{(g)} \\ \boldsymbol{\delta}_{2}^{(g)} \\ \vdots \\ \boldsymbol{\delta}_{U_{g}}^{(g)} \end{bmatrix}}}.
\label{Main_eq}
\end{align}
Let $\mathbf{w}_{\mathrm{BB}}^{(g)} = \left[\left(\mathbf{w}_{\mathrm{BB},1}^{(g)}\right)^T, \left(\mathbf{w}_{\mathrm{BB},2}^{(g)}\right)^T,\ldots,\left(\mathbf{w}_{\mathrm{BB},U_g}^{(g)}\right)^T\right]^T$ denote the concatenated baseband combiner of all the users for the $g$th group. The signal $\check{r}_u^{(g)} = \left(\mathbf{w}_{\mathrm{BB},u}^{(g)}\right)^H \hat{\mathbf{r}}_{u}^{(g)}$ received by the $g$th group after baseband RC is given by
\begin{align}\label{McastInt}
\check{r}_u^{(g)} =& \left(\mathbf{w}_{\mathrm{BB}}^{(g)}\right)^H\mathbf{H}^{(g)}_{\mathrm{eff}}\mathbf{f}_{\mathrm{BB}}^{(g)} x^{(g)} + \left(\mathbf{w}_{\mathrm{BB}}^{(g)}\right)^H\sum_{\substack{l=1\\ l \neq i}}^{G}\mathbf{H}^{(g)}_{\mathrm{eff}}\mathbf{f}_{\mathrm{BB}}^{(l)} x^{(l)} \nonumber \\ +& \check{{\boldsymbol{\delta}}}^{(g)}.
\end{align}
The proposed OSHB-M algorithm employs the ZF constraint $\mathbf{G}^{(g)}\mathbf{f}_{\mathrm{BB}}^{(g)}=\mbox{0}$ for cancelling the interference denoted by the second term in \eqref{McastInt}, where the matrix $\mathbf{G}^{(g)} \in \mathbb{C}^{(\sum_{j=1}^{g-1}U_j) \times \sum_{m=1}^{M}N_{\mathrm{RF},m}}$ is defined as $\mathbf{G}^{(g)} = \left[\left(\mathbf{F}^{(1)}\right)^T, \left(\mathbf{F}^{(2)}\right)^T, \ldots, \left(\mathbf{F}^{(g-1)}\right)^T\right]^T$, with

\begin{align}
\mathbf{F}^{j}=\begin{bmatrix} \left(\mathbf{w}_{\mathrm{BB},1}^{(g)}\right)^{H} \mathbf{H}^{(j)}_{\mathrm{eff},1}\\ \left(\mathbf{w}_{\mathrm{BB},2}^{(g)}\right)^{H} \mathbf{H}^{(j)}_{\mathrm{eff},2}\\ \vdots\\ \left(\mathbf{w}_{\mathrm{BB},U_{j}}^{(g)}\right)^{H} \mathbf{H}^{(j)}_{\mathrm{eff},U_{j}} \end{bmatrix} \in \mathbb{C}^{U_j \times MN_\mathrm{RF},m}.
\end{align}
Hence, the baseband TPC constructed for the $g$th group is given by $\mathbf{f}_{\mathrm{BB}}^{(g)} = {(\mathbf{G}^{(g)})}^{\perp} \boldsymbol{\lambda}^{(g)}$, where $\boldsymbol{\lambda}^{(g)} \in \mathbb{C}^{\left(\sum_{m=1}^{M}N_{\mathrm{RF},m} - \sum_{j=1}^{g-1}U_j\right) \times 1}$, and the columns of the matrix ${(\mathbf{G}^{(g)})}^{\perp} \in \mathbb{C}^{\sum_{m=1}^{M}N_{\mathrm{RF},m} \times (MN_{\mathrm{RF},m} - \sum_{j=1}^{g-1}U_j)}$ constitute an orthonormal basis for the null-space of the matrix $\mathbf{G}^{(g)}$, and can be obtained as described in Section IV for the matrix $\mathbf{G}_{u-1}^{\perp}$. Next, we denote the interference-plus-noise part $\boldsymbol{\tau}^{(g)}$ in  (\ref{Main_eq}) as $\boldsymbol{\tau}^{(g)} = \sum_{l=1}^{g-1} \mathbf{H}^{(g)}_{\mathrm{eff}} \mathbf{f}_{\mathrm{BB}}^{(l)} + \boldsymbol{\delta}^{(g)}$. The covariance matrix $\mathbf{C}_{\tau}^{(g)} = \mathrm{E}[\boldsymbol{\tau}^{(g)} {(\boldsymbol{\tau}^{(g)})}^H] \in \mathbb{C}^{U_{g}N_{\mathrm{RF},u}\times U_{g}N_{\mathrm{RF},u}}$ of $\boldsymbol{\tau}^{(g)}$ for the $g$th multicast group can be written as
\begin{equation*}
\mathbf{C}_{\tau}^{(g)} = \sum_{l=1}^{g-1}\mathbf{H}^{(g)}_{\mathrm{eff}}\left(\mathbf{f}_{\mathrm{BB}}^{(l)}\left({\mathbf{f}_{\mathrm{BB}}^{(l)}}\right)^{H}\right) \left(\mathbf{H}^{(g)}_{\mathrm{eff}}\right)^{H}+\sigma_\delta^2 \mathbf{I}_{U_{g}N_{\mathrm{RF},u}}.
\end{equation*}
Therefore, the optimal baseband TPC $\mathbf{f}_{\mathrm{BB}}^{(g)}$ and RC $\mathbf{w}_{\mathrm{BB}}^{(g)}$ conditioned for the $g$th multicast group are given as $\mathbf{f}_{\mathrm{BB}}^{(g)} = {(\mathbf{G}^{(g)})}^{\perp} \boldsymbol{\lambda}^{(g)}$ and $\mathbf{w}_{\mathrm{BB}}^{(g)} = {\left({\left(\mathbf{C}_{\tau}^{(g)}\right)}^{-({1}/{2})}\right)}^H \tilde{\mathbf{w}}_{\mathrm{BB}}^{(g)}$, where $\tilde{\mathbf{w}}_{\mathrm{BB}}^{(g)}$ and $\boldsymbol{\lambda}^{(g)}$ represent the principal left and right singular vectors, respectively, of the matrix $\tilde{\mathbf{H}}^{(g)}_{\mathrm{eff}} = {(\mathbf{C}_{\tau}^{(g)})}^{-({1}/{2})}\mathbf{H}^{(g)}_{\mathrm{eff}}{(\mathbf{G}^{(g)})}^{\perp} \in \mathbb{C}^{U_g N_{\mathrm{RF},u} \times (MN_{\mathrm{RF},m} - \sum_{j=1}^{g-1}U_j)}$. This result follows from our analysis similar to that provided for the unicast scenario described in Appendix A. Finally, it can be readily observed that the proposed OSHB-U and OSHB-M schemes can be implemented at the complexity order of $\mathcal{O}\left(\left(\sum_{m=1}^{M}N_{\mathrm{RF},m}\right)^3\right)$. This is due to the fact that these schemes are based on the singular value decomposition of the matrices $\mathbf{M}_{u-1}$ and $\mathbf{G}^{(g)}$. Note that while deriving the multicast beamforming solutions we assumed that the multicast users are in the same geographical location. However, if the users that request the same data are in different geographical regions, it is more efficient to transmit separately, i.e., unicast, to each user \cite{sadeghi2018joint}.
\section{Hybrid precoder design using the Bayesian learning (BL)-based approach }
One can note that in the previous sections, the designs of the RF TPCs and RCs require the knowledge of the dominant array response vectors of the APs and users, respectively, which eventually necessitates the full knowledge of the AoAs/AoDs of the multipath components and their complex-valued path gains. However, in practice, it is challenging to acquire accurate estimates of these quantities, which motivates us to develop a scheme that can efficiently design the hybrid TPCs and RCs without requiring the knowledge of the full channel state information (CSI).

To this end, in this section, we jointly design the baseband and RF TPCs and RCs for a unicast scenario. However, a similar approach can be readily developed for a general multicast scenario. The proposed joint design first obtains the optimal digital baseband TPC and RC $\mathbf{f}_{\mathrm{opt}}$ and $\mathbf{w}_{\mathrm{opt}}$, respectively, assuming a fully digital architecture. This step only requires the knowledge of the mmWave MIMO channel matrices $\mathbf{H}_{u,m}$, rather than their component AoAs/AoDs and path gains. Subsequently, a BL-based algorithm decomposes the fully digital precoders/combiners $\mathbf{f}_{\mathrm{opt}}$ and $\mathbf{w}_{\mathrm{opt}}$ into their RF and baseband components. The signal $\mathbf{r}_u \in \mathbb{C}^{N_R \times 1}$ received by the $u$th user can be expressed as
\begin{align}
\mathbf{r}_u = \sum_{m=1}^{M}\mathbf{H}_{u,m}\mathbf{f}_{u,m}x_u + \sum_{\substack{i=1\\ i \neq u}}^{U} \sum_{m=1}^{M}\mathbf{H}_{u,m}\mathbf{f}_{i,m}x_i + \boldsymbol{\delta}_{u}.
\end{align}
Let the concatenated fully digital TPC and channel matrix for the $u$th user be defined as $\mathbf{f}_{u} = [\mathbf{f}_{u,1}^{T}, \mathbf{f}_{u,2}^{T}, \ldots, \mathbf{f}_{u,M}^{T}]^{T}$, and 
$\mathbf{H}_{u} = \left[\mathbf{H}_{u,1} \mathbf{H}_{u,2} \cdots \mathbf{H}_{,u,M}\right]$, respectively. Employing the fully digital RC $\mathbf{w}_{u} \in \mathbb{C}^{N_R \times 1}$, the signal $\tilde{r}_u = \mathbf{w}_{u}^H\mathbf{r}_u$ received by the $u$th user can be written as
\begin{align}
\tilde{r}_u =  \mathbf{w}_{u}^H\mathbf{H}_{u}\mathbf{f}_{u}x_u + \sum_{\substack{i=1\\ i \neq u}}^{U} \mathbf{w}_{u}^H\mathbf{H}_{u}\mathbf{f}_{i}x_i + \underset{\tilde{\boldsymbol{\delta}}_u}{\underbrace{\mathbf{w}_{u}^H \boldsymbol{\delta}_u}}.
\end{align}
The SINR of the $u$th user is given by
\begin{align}
\text{SINR}_u = \frac{\left|\mathbf{w}_{u}^H\mathbf{H}_{u}\mathbf{f}_{u}\right|^2}{\sum_{i=1,i \neq u}^{U}{\left|\mathbf{w}_{u}^H\mathbf{H}_{u}\mathbf{f}_{i}\right|^2}+ \sigma_{\tilde{\delta}}^2}.
\end{align}
Upon employing the proposed OSHB-U algorithm, the optimal fully digital TPC and RC are obtained as $\mathbf{f}_{u}=\mathbf{A}_{u-1}^{\perp} \boldsymbol{\nu}_{u}$ and $\mathbf{w}_{u} = {(\mathbf{R}_{\tau_{u}}^{-(1/2)})}^H \tilde{\mathbf{w}}_{u}$ respectively, where $\boldsymbol{\nu}_{u}$ and $\tilde{\mathbf{w}}_{u}$ are the principal right and left singular vectors, respectively, of the matrix $\mathbf{R}_{\tau_{u}}^{-({1}/{2})}\mathbf{H}_{u}\mathbf{A}_{u-1}^{\perp} \in \mathbb{R}^{N_{R} \times (NN_{T} - u+1)}$. The columns of the matrix $\mathbf{A}_{u-1}^{\perp}\in\mathbb{C}^{MN_T \times \left(MN_T -u+1\right)}$ represent a basis orthonormal to the null-space of the matrix $\mathbf{A}_{u-1}=\left[\left(\mathbf{w}_{1}^H\mathbf{H}_{1}\right)^T,\ldots, \left(\mathbf{w}_{u-1}^H\mathbf{H}_{u-1}\right)^T\right]^T \in\mathbb{C}^{\left(u-1\right) \times MN_T}$. Furthermore, the concatenated fully digital TPC $\mathbf{F}_{\mathrm{opt}} \in \mathbb{C}^{MN_T \times U}$ corresponding to all users is given by $\mathbf{F}_{\mathrm{opt}} = \left[\mathbf{f}_{1}, \mathbf{f}_{2}, \ldots, \mathbf{f}_{U} \right]$. Now, the decomposition problem of jointly designing the baseband TPC $\mathbf{F}_{\mathrm{BB}} \in \mathbb{C}^{N_{\sum_{m=1}^{M}\mathrm{RF},m} \times U}$ and RF TPC $\mathbf{F}_{\mathrm{RF}} \in \mathbb{C}^{MN_{T} \times \sum_{m=1}^{M} N_{\mathrm{RF},m}}$ can be formulated as
\begin{align}\label{SBL1}
& \!\!\!\!\!\!\!\!\!\!\!\!\! \left(\mathbf{F}_{\mathrm{BB}}^{*}, \mathbf{F}_{\mathrm{RF}}^{*}\right)= \underset{\mathbf{F}_{\mathrm{BB}}, \mathbf{F}_{\mathrm{RF}}}{\arg \ \ \  \min}\left\|\mathbf{F}_{\mathrm{opt}} -\mathbf{F}_{\mathrm{RF}} \mathbf{F}_{\mathrm{BB}}\right\|_{\mathrm{F}}^2\nonumber \\ 
& \hspace{2cm} \text { s.t. } \ \ \ \ \vert \mathbf{F}_{\mathrm{RF}}(i,j)\vert = \frac{1}{\sqrt{N_T}},
\end{align}
where $\mathbf{F}_\mathrm{BB} = \left[\mathbf{f}_{\mathrm{BB},1}, \mathbf{f}_{\mathrm{BB},2},\ldots, \mathbf{f}_{\mathrm{BB},U}\right]$ denotes the concatenated baseband TPC and $\mathbf{F}_{\mathrm{RF}}=\mathrm{blkdiag}\left(\mathbf{F}_{\mathrm{RF},1},\ldots,\mathbf{F}_{\mathrm{RF},m}\right)$ represents the block-diagonal matrix of RF TPCs corresponding to all the APs. The constant magnitude constraint in \eqref{SBL1} is due to the analog phase shifter based architecture, which results in a non-convex optimization problem. Towards solving this, one can first define a dictionary matrix of the feasible transmit array response vectors as $\mathbf{G}_{T} = \left[\mathbf{a}_{T}\left(\phi_{1}\right), \mathbf{a}_{T}\left(\phi_{2}\right), \cdots, \mathbf{a}_{T}\left(\phi_{S}\right)\right] \in \mathbb{C}^{N_{T} \times S}$, where the set of AoDs $\{\phi_{s},\ \forall 1\leq s\leq S\}$ spans the angular range $[0,\ \pi]$, obeying $\displaystyle \cos(\phi_{s})=\frac{2}{S}(s-1)-1$, and the quantity $S$ represents the angular grid size \cite{heath2016overview}. Note that the elements of the dictionary matrix satisfy the constant magnitude constraint of \eqref{SBL1}. Hence, the columns of the RF TPC $\mathbf{F}_{\mathrm{RF}}$ can be readily chosen from the dictionary matrix $\mathbf{G}_T$.
The equivalent problem for hybrid precoder design in the CFMU mmWave MIMO system can be formulated as
\begin{equation}\label{SBLmain}
\begin{array}{c} 
\underset{\tilde{\mathbf{F}}_{\mathrm{BB}}}{\arg \min}\left\|\mathbf{F}_{\mathrm{opt}}-\mathbf{G}_{T} \tilde{\mathbf{F}}_{\mathrm{BB}}\right\|_{\mathrm{F}} \\ 
\hspace{1.2cm}\mathrm{ s.t. } \hspace{0.5cm} \left\|\operatorname{diag}\left(\tilde{\mathbf{F}}_{\mathrm{BB}} \tilde{\mathbf{F}}_{\mathrm{BB}}^{H}\right)\right\|_{0} = \sum_{m=1}^{M} N_{\mathrm{RF},m},
\end{array}
\end{equation}
where $\tilde{\mathbf{F}}_{\mathrm{BB}} \in \mathbb{C}^{S \times U}$ denotes the intermediate baseband TPC. The constraint in \eqref{SBLmain} arises due to the fact that there can only be $\sum_{m=1}^{M} N_{\mathrm{RF},m}$ non-zero rows in the matrix $\tilde{\mathbf{F}}_{\mathrm{BB}}$, since there are only $N_{\mathrm{RF},m}$ RF chains at each AP. This leads to a simultaneous sparse structure in the matrix $\tilde{\mathbf{F}}_{\mathrm{BB}}$. The procedure of obtaining the proposed BL-based hybrid TPC corresponding to the optimization problem \eqref{SBLmain} is discussed next.

The BL method proposed for our hybrid TPC design assigns the parameterized Gaussian prior $p\left(\displaystyle \tilde{\mathbf{F}}_{\mathrm{BB}};{\boldsymbol{\Gamma}}\right)$ to the baseband TPC matrix $\tilde{\mathbf{F}}_{\mathrm{BB}}$, where we have
\begin{align}
p\left(\displaystyle \tilde{\mathbf{F}}_{\mathrm{BB}};{\boldsymbol{\Gamma}}\right)=&\, \prod_{i=1}^{S}p\left(\tilde{\mathbf{F}}_{\mathrm{BB}}(i,:);\gamma_{i}\right) \nonumber \\ =&\, \prod_{i=1}^{S} \frac{1}{\pi\gamma_i} \exp\left(- \frac{\norm{\tilde{\mathbf{F}}_{\mathrm{BB}}(i,:)}^2}{\gamma_i}\right),
\end{align}
and $\boldsymbol{\Gamma}=\mathrm{diag}\left(\gamma_{1},\ldots,\gamma_{S}\right) \in \mathbb{R}^{S \times S}$  denotes the diagonal matrix of hyper-parameters. One can observe that the $i$th row of the matrix $\tilde{\mathbf{F}}_{\mathrm{BB}}$ is assigned the hyperparameter $\gamma_i$, which enforces the row sparsity, as given in the constraint \eqref{SBLmain}. The MMSE estimator of the matrix $\tilde{\mathbf{F}}_{\mathrm{BB}}$, represented by the matrix $\boldsymbol{\Phi} \in \mathbb{C}^{S \times U}$ and the associated covariance matrix ${\boldsymbol{\Pi}}\in \mathbb{C}^{S \times S}$ are given as
\begin{equation}
\boldsymbol{\Phi} =\frac{1}{\sigma_{e}^{2}}\boldsymbol{\Pi}\mathbf{G}_{T}^{H}\mathbf{F}_{\mathrm{opt},n} \ \ \text{and} \ \ \displaystyle \mathbf{\Pi}=\left(\frac{1}{\sigma_{e}^{2}}\mathbf{G}_{T}^{H}\mathbf{G}_{T}+\mathbf{\Gamma}^{-1}\right)^{-1},
\label{eq. mu and sigma}
\end{equation}
where $\sigma_{e}^2$ denotes the variance of the approximation error. It can be observed that the MMSE estimate $\boldsymbol{\Phi}$ depends on the hyperparameter matrix $\mathbf{\Gamma}$. Furthermore, as $\gamma_i \rightarrow 0$, the $i$th row of the matrix $\tilde{\mathbf{F}}_{\mathrm{BB}}$ denoted by $\tilde{\mathbf{F}}_{\mathrm{BB}}(i,:)\rightarrow0$. Thus, the estimation of $\tilde{\mathbf{F}}_{\mathrm{BB}}$ is equivalent to the estimation of the associated hyperparameter vector $\boldsymbol{\gamma}=[\gamma_{1},\ldots,\gamma_{S}]^{T}$. To find the maximal likelihood estimate of $\boldsymbol{\gamma}$, the Bayesian evidence $p(\mathbf{F}_{\mathrm{opt}};\boldsymbol{\Gamma)}$ can be maximized. However, the pertinent optimization problem becomes non-convex, and it is hence difficult to solve. Therefore, the BL technique harnessed for hybrid TPC design in CFMU systems relies on a low-complexity iterative expectation-maximization (EM) framework, which guarantees likelihood maximization in each iteration, and convergence to a local optimum.

To begin with, let $\hat{\boldsymbol{\Gamma}}^{(k-1)} $ denote the estimate of the hyperparameter matrix $\boldsymbol{\Gamma}$ obtained from the $(k-1)$st iteration. The EM procedure involves two key steps. In the first step, known as the expectation step (E-step), we evaluate the average log-likelihood function $\mathcal{L}\left(\boldsymbol{\Gamma}\mid\hat{\boldsymbol{\Gamma}}^{(k-1)}\right)$ of the hyper-parameters as follows
\begin{align*}
\mathcal{L}\left(\boldsymbol{\Gamma}\mid\hat{\boldsymbol{\Gamma}}^{(k-1)}\right) = \mathbb{E}_{\tilde{{\bf F}}_{\mathrm{BB}}\mid{\bf F}_{\mathrm{opt}};\hat{\boldsymbol{\Gamma}}^{(k-1)}}\Bigg \lbrace\log p\left({\bf F}_{\mathrm{opt}},\tilde{{\bf F}}_{\mathrm{BB}};\boldsymbol{\Gamma}\right)\Bigg \rbrace.
\end{align*}
The second step, which is known as the maximization step (M-step), maximizes the average log-likelihood computed above with respect to the hyperparameter vector $\boldsymbol{\gamma}$ as
\begin{align}
\hat{\boldsymbol{\gamma}}^{(k)}=\arg \max _{\boldsymbol{\gamma}} \mathbb{E} \Bigg \lbrace\log p\left(\mathbf{F}_{\mathrm{opt}} \mid \tilde{\mathbf{F}}_{\mathrm{BB}}\right) + \log p\left(\tilde{\mathbf{F}}_{\mathrm{BB}} ; \boldsymbol{\Gamma}\right)\Bigg \rbrace. \label{eqHyp}
\end{align}
It can be readily observed from \eqref{eqHyp} that the expression $p\left(\mathbf{F}_{\mathrm{opt}} \mid \tilde{\mathbf{F}}_{\mathrm{BB}}\right)$ is independent of the hyperparameter $\boldsymbol{\gamma}$ and can be disregarded in the successive M-step. Therefore, the equivalent optimization problem is formulated as
\begin{align}\label{estimate_gamma}
\hat{\boldsymbol{\gamma}}^{(k)} =&\, \arg \max _{\boldsymbol{\gamma}} \mathbb{E}_{\tilde{\mathbf{F}}_{\mathrm{BB}} \mid \mathbf{F}_{\mathrm{opt}},\hat{\boldsymbol{\Gamma}}^{(k-1)}}\left\{\log p\left(\tilde{\mathbf{F}}_{\mathrm{BB}} ; \boldsymbol{\Gamma}\right)\right\}\nonumber \\
\equiv &\, \arg \max _{\boldsymbol{\gamma}} \sum_{i=1}^{S}\left[ -\log \left( \gamma_{i}\right)-\frac{\mathbb{E}\left(\norm{\tilde{\mathbf{F}}_{\mathrm{BB}}(i,:)}^{2}_{2}\right)}{\gamma_{i}}\right], \nonumber \\
 \equiv &\, \arg \max _{\boldsymbol{\gamma}} \sum_{i=1}^{S} -\log \left( \gamma_{i}\right)-\frac{\norm{\boldsymbol{\Phi}^{(k)}(i,:)}^{2} + U \boldsymbol{\Pi}_{(i, i)}^{(k)}}{\gamma_{i}},
\end{align}
where $\boldsymbol{\Phi}^{(k)}$ and $\boldsymbol{\Pi}^{(k)}$ are obtained from (\ref{eq. mu and sigma}) by setting $\boldsymbol{\Gamma} = \hat{\boldsymbol{\Gamma}}^{(k-1)}$. The optimal value of $\hat{\gamma}_{i}^{(k)}$  can be evaluated by equating the gradient of the objective function in \eqref{estimate_gamma} with respect to $\boldsymbol{\gamma}$ to zero, which yields the update equation for each hyperparameter as
\begin{equation*}
\displaystyle \hat{\gamma}_{i}^{(k)}= \frac{1}{U} \norm{\boldsymbol{\Phi}^{(k)}(:,i)}^2+\boldsymbol{\Pi}^{(k)}_{(i,i)}.
\end{equation*}
Upon convergence of the EM procedure, the RF and baseband TPCs are obtained as follows.
Let $\mathcal{S}$ denote the set of indices for the $\left(\sum_{m=1}^{M}N_{\mathrm{RF},m}\right)$ hyper-parameters having the largest magnitudes.  The concatenated optimal baseband TPC matrix $\mathbf{F}_{\mathrm{BB}}^{*}$ can be extracted from $\tilde{\mathbf{F}}_{\mathrm{BB}}$ as
\begin{align}\label{FbbBL}
\mathbf{F}_{\mathrm{BB}}^{*} = \tilde{\mathbf{F}}_{\mathrm{BB}}\left(\mathcal{S},:\right).
\end{align}
Similarly, the optimal RF TPC $\mathbf{F}_{\mathrm{RF}}^{*}$ can be obtained from the dictionary matrix of the transmit array response $\mathbf{G}_T$ as
\begin{align}\label{FrfBL}
\mathbf{F}_{\mathrm{RF}}^{*} = \mathbf{G}_{T}\left(:,\mathcal{S} \right).
\end{align}
Algorithm 2 presents the step-by-step procedure of the BL-based technique for designing the hybrid TPC. One can follow a similar procedure to obtain the digital baseband and RF RCs from the fully digital design. The complexity analysis of the proposed two-stage distributed hybrid beamformer design is summarized next. Due to lack of space, the detailed derivations for the computational complexities of Algorithms 1 and 2 have been relegated to our technical report \cite{TechReport}. It can be observed that the complexity of Algorithm-1, which designs the fully-digital transmit precoder (TPC), is $\mathcal{O}\left(\left(U-1 \right)^2 MN_T\right)$. Next, the fully-digital TPC is decomposed into its constituent radio frequency (RF) and baseband (BB) precoders using Algorithm-2. This step incurs a complexity of order $\mathcal{O}\left(S^3\right)$, which can be attributed to the matrix inversion of size $\left[S \times S \right]$ in Eq. 29. Since, $S >> \left(U-1 \right)^2MN_T$, the overall complexity of the coordinated hybrid TPC can be closely approximated by $\mathcal{O}\left(S^3\right)$. Furthermore, the global convergence of the BL algorithm's cost function is obtained at the sparsest solution, which guarantees the sparsest representation of the digital precoder \cite{wipf2004sparse}. Further, as a benefit of the EM-algorithm, the proposed BL technique is guaranteed to converge to a fixed point of the log-likelihood function, regardless of its initialization. This leads to the robust performance of the BL algorithm and makes it ideally suited for hybrid precoder/ combiner design in mmWave MIMO networks.
\begin{algorithm}\label{BLtechnique}
\SetAlgoLined
\textbf{Input:} Concatenated optimal digital precoder matrix $\mathbf{F}_{\mathrm{opt}}$, dictionary matrix $\mathbf{G}_T$, RF chains $\sum_{m=1}^{M}N_{\mathrm{RF},m}$, variance of approximation error $\sigma_e^2$, stopping parameters $\epsilon$ and $k_{\mathrm{max}}$\;
\textbf{Initialization:} $\hat{\gamma}_i^{(0)} = 1$, $\forall 1 \leq i \leq S \rightarrow \hat{\boldsymbol{\Gamma}}^{(0)} = \mathbf{I}_S$. Set counter $k = 0$ and $\boldsymbol{\Gamma}^{(-1)} = \mathbf{0}$\;
 \While{$\left(\norm{\hat{\boldsymbol{\gamma}}^{(k)} - \hat{\boldsymbol{\gamma}}^{(k-1)}}^2 > \epsilon \ \mathrm{and} \ k < k_{\mathrm{max}}\right)$}{
  \textbf{E-step:} Evaluate the \textit{a posteriori} covariance and mean $$\boldsymbol{\Gamma}^{(k)} = \left(\frac{1}{\sigma^2_e}\mathbf{G}_T^H \mathbf{G}_T + \left(\hat{\boldsymbol{\Gamma}}^{(k-1)}\right)^{-1}\right)^{-1}$$ 
  $$\!\!\!\!\!\!\!\!\!\!\!\!\!\!\!\!\!\!\!\!\!\!\!\!\!\!\!\!\!\!\!\!\!\!\!\!\!\!\! \tilde{\mathbf{F}}_{\mathrm{BB}}^{(k)} = \frac{1}{\sigma^2_e}\boldsymbol{\Pi}^{(k)}\mathbf{G}_T^H\mathbf{F}_{\mathrm{opt}}$$ \
    \textbf{M-step:} Evaluate the hyperparameter estimate\
   $$\!\!\!\!\!\!\!\!\!\!\!\!\!\!\!\!\!\!\!\!\!\!\!\!\!\!\!\!\!\!\!\!\!\!\!\!\!\!\!\!\!\!\!\!\!\!\!\!\!\!\!\!\!\! \textbf{for} \ i = 1,\ldots, S$$
   $$\hspace{1.2cm} \hat{\gamma}_{i}^{(k)} = \boldsymbol{\Pi}^{(k)}_{(i,i)} + \frac{1}{U} \sum_{j=1}^{U}\left|\tilde{\mathbf{F}}_{\mathrm{BB},n}(i,j)\right|^2 $$
   \ \ \ \ \ \ \ \textbf{end}\
 } \textbf{Output:} Obtain $\mathbf{F}_{\mathrm{BB}}^{*}$ and $\mathbf{F}_{\mathrm{RF}}^{*}$ using \eqref{FbbBL} and \eqref{FrfBL}.
 \caption{BL-based hybrid TPC design}
\end{algorithm}
\section{Successive Multiuser Uplink Hybrid Beamforming (SCUHBF)}
Consider now the cooperative CFMU mmWave MIMO uplink system with $U$ users and $M$ cooperative APs. The $u$th user has $N_R$ TAs and $N_{\mathrm{RF},u}$ RF chains, whereas the $m$th AP is equipped with $N_T$ RAs and $N_{\mathrm{RF},m}$ RF chains. The symbol transmitted by user $u$ can be expressed as $ \mathbf{s}_u = \mathbf{F}_{\mathrm{RF},u}\mathbf{f}_{\mathrm{BB},u}x_u.$
The signal received at the $m$th AP can be expressed as
\begin{align*}
\mathbf{r}_m = \sum_{u=1}^{U}\mathbf{H}_{u,m}\mathbf{s}_u+ \boldsymbol{\delta}_m = \sum_{u=1}^{U}\mathbf{H}_{u,m}\mathbf{F}_{\mathrm{RF},u}\mathbf{f}_{\mathrm{BB},u}x_u + \boldsymbol{\delta}_m,
\end{align*}
where $\mathbf{F}_{\mathrm{RF},u}\in \mathbb{C}^{N_R \times N_{\mathrm{RF},u}}$ represents the RF TPC and $\mathbf{f}_{\mathrm{BB},u}\in \mathbb{C}^{N_{\mathrm{RF},u} \times 1}$ denotes the baseband TPC of the user $u$. Furthermore, $\mathbf{H}_{u,m} \in \mathbb{C}^{N_T \times N_R}$ denotes the channel matrix between the $m$th AP and the user $u$. Therefore, the stacked signal $\mathbf{r} = \left[\mathbf{r}_1^H, \mathbf{r}_2^H,\ldots, \mathbf{r}_M^H\right]^H \in \mathbb{C}^{MN_T \times 1} $ received at all the cooperating APs can be written as
\begin{align}
\mathbf{r} = \sum_{u=1}^{U}\mathbf{H}_u\mathbf{F}_\mathrm{{RF},u}\mathbf{f}_\mathrm{{BB},u}x_u + \boldsymbol{\delta},
\end{align}
where $\mathbf{H}_u = {[\mathbf{H}_{u,1}^H, \mathbf{H}_{u,2}^H, \ldots, \mathbf{H}_{u,M}^H]}^H \in \mathbb{C}^{MN_T \times N_R}$ denotes the aggregate channel matrix between all the APs and the $u$th user. Also, ${\boldsymbol{\delta}} \in \mathbb{C}^{MN_T \times 1}$ denotes the complex AWGN with covariance matrix $\mathop{\mathbb{E}}[\boldsymbol{\delta} \boldsymbol{\delta}^H] = \sigma_{\delta}^2 \mathbf{I}_{MN_T}$. The estimated received symbol corresponding to user $u$ after employing the hybrid RC $\mathbf{w}_u = \mathbf{W}_{\rm RF}\mathbf{w}_{\mathrm{BB},u}$ at the cooperative APs is given by
\begin{align}\label{UplinkEstimated}
\hat{x}_u =& \mathbf{w}^H_u \mathbf{r} = \mathbf{w}^H_u \mathbf{H}_u \mathbf{s}_u  + \sum_{i=1,i \neq u}^{U} \mathbf{w}^H_u \mathbf{H}_i \mathbf{s}_i  + \underset{\tilde{\boldsymbol{\delta}}}{\underbrace{\mathbf{w}_{u}^H \boldsymbol{\delta}}}\nonumber\\ 
=& \mathbf{w}^H_{\mathrm{BB},u} \mathbf{W}^H_{\mathrm{RF}} \mathbf{H}_u \mathbf{F}_{\mathrm{RF},u}\mathbf{f}_{\mathrm{BB},u}x_u \nonumber \\ +& \sum_{i=1,i \neq u}^{U} \mathbf{w}^H_{\mathrm{BB},u} \mathbf{W}^H_{\rm RF} \mathbf{H}_i \mathbf{F}_{\mathrm{RF},i}\mathbf{f}_{\mathrm{BB},i}x_i  + \tilde{{\boldsymbol{\delta}}}.
\end{align}
Let us define the effective channel matrix $\mathbf{H}_{\mathrm{eff},u}$ between the user $u$ and the APs as $\mathbf{H}_{\mathrm{eff},u} = \mathbf{W}^H_{\rm RF} \mathbf{H}_u \mathbf{F}_{\mathrm{RF},u} \in \mathbb{C}^{MN_{\mathrm{RF},m} \times N_{\mathrm{RF},u}}$. From \eqref{UplinkEstimated}, the effective system model for the estimated symbol $\hat{x}_u$ is given by
\begin{equation}
\hat{x}_u = \mathbf{w}^H_{\mathrm{BB},u} \mathbf{H}_{{\rm eff},u}\mathbf{f}_\mathrm{{BB},u}x_u + \sum_{i=1,i \neq u}^{U}\mathbf{w}^H_{\mathrm{BB},u} \mathbf{H}_{{\rm eff},i}\mathbf{f}_{\mathrm{BB},i}x_i + \tilde{{\boldsymbol{\delta}}}.
\label{Eq7MRC}
\end{equation}

\begin{figure*}[t]
\centering
\subfloat[]{\includegraphics[width=0.45\linewidth]{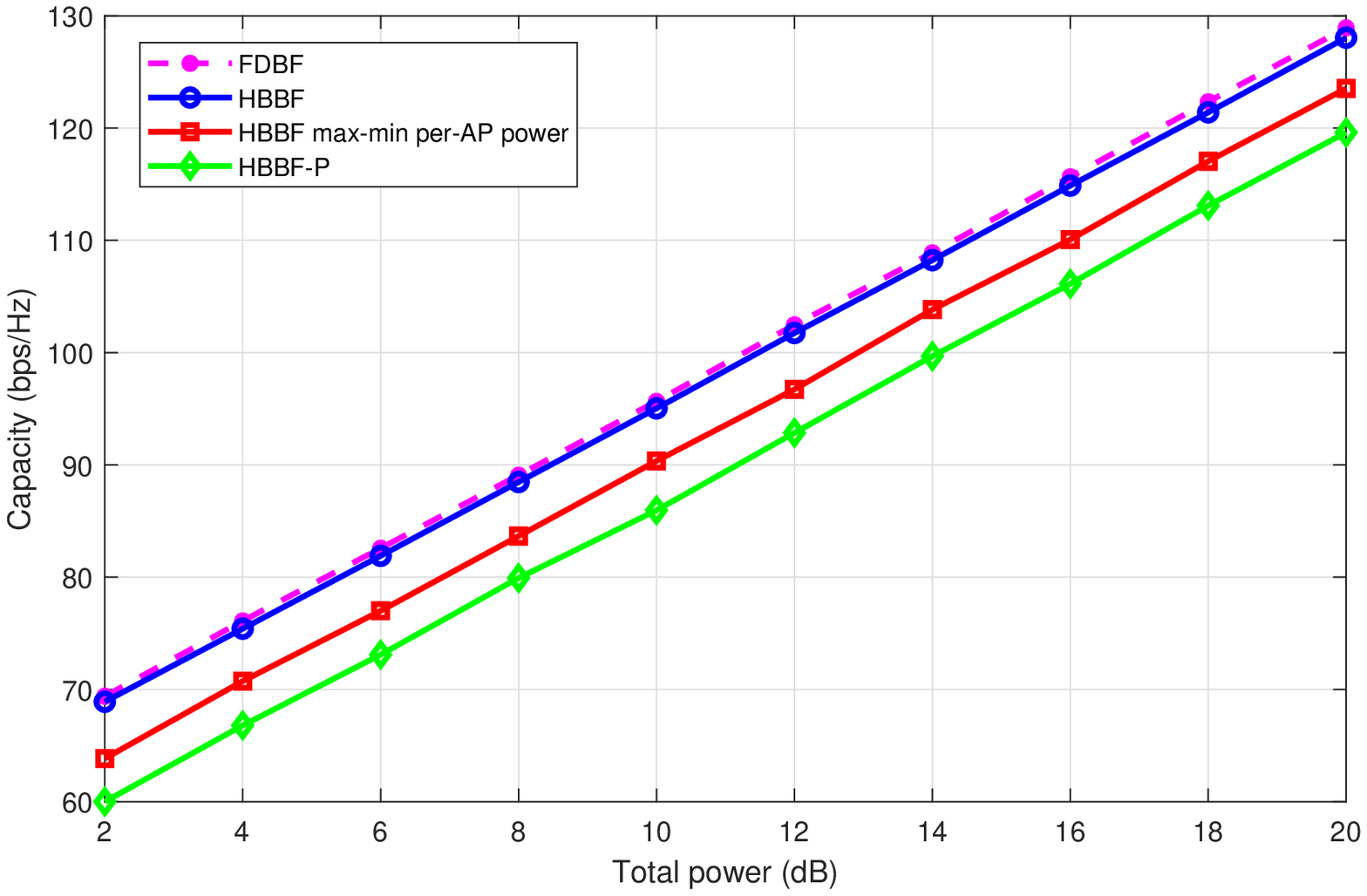}\label{Broadcast}}
\subfloat[]{\includegraphics[width=0.45\linewidth]{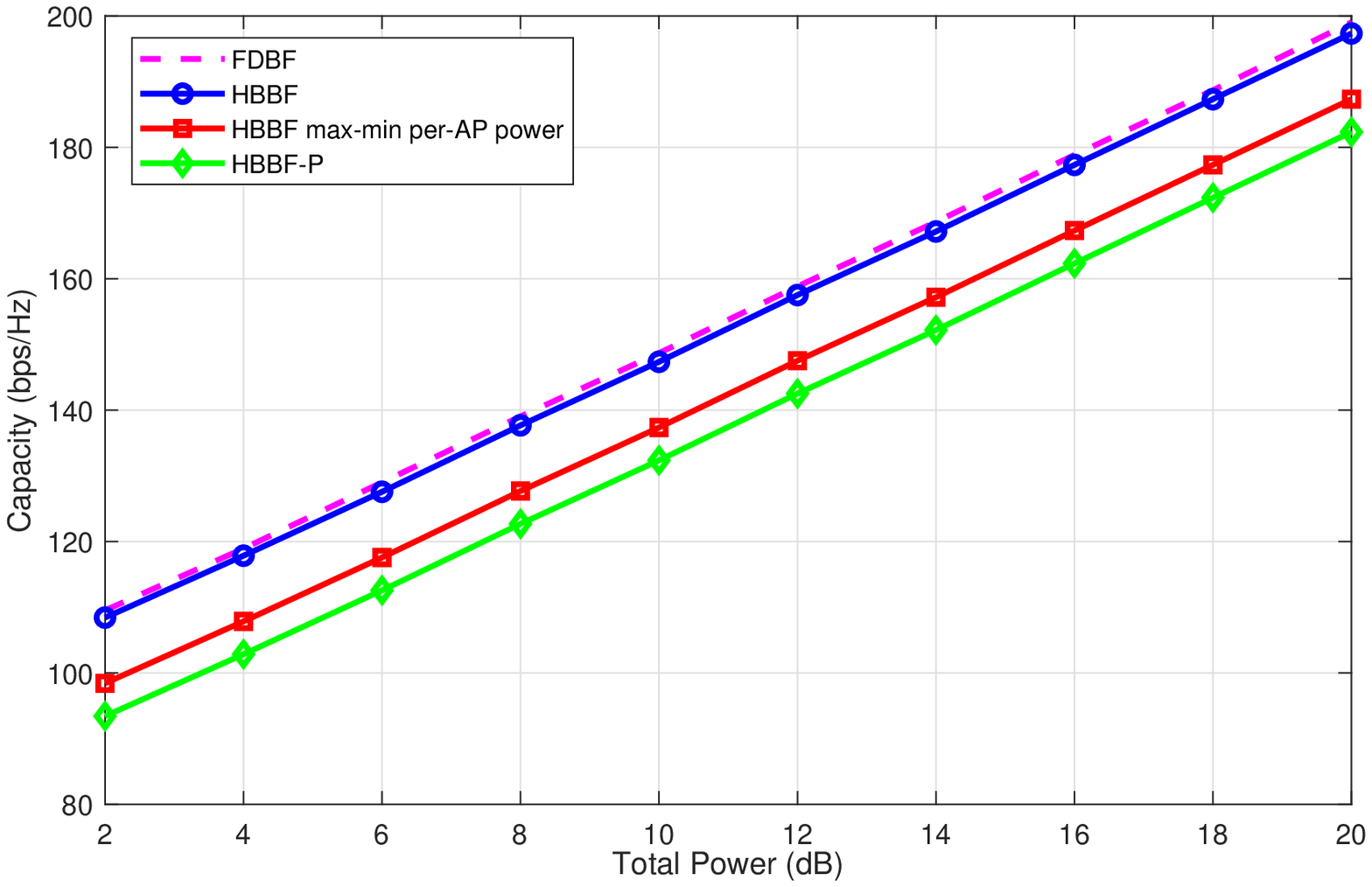}\label{BroadcastLargeAntenna}}
\caption{Capacity comparison of various schemes for a broadcast scenario with(a) $M = 4$, $N_T = 16$, $N_R = 8$, $N_{\mathrm{RF},m} = 4$, and $U = 10$, (b) $M = 4$, $N_T = 64$, $N_R = 8$, $N_{\mathrm{RF},m} = 4$, and $U = 15$. \vspace{-10pt}}\label{Broadcast_Main_Fig}
\end{figure*}
The baseband RC designed for the $u$th user is given by $\mathbf{w}_{\mathrm{BB},u} = \mathbf{H}_{{\rm eff},u}\mathbf{f}_{\mathrm{BB},u}$. It can be observed from \eqref{Eq7MRC} that the term $\mathbf{f}_{\mathrm{BB},u}^H \left(\mathbf{H}_{\mathrm{eff},u}^H \mathbf{H}_{\mathrm{eff},u}\right) \mathbf{f}_{\mathrm{BB},u}$ represents the signal power of the $u$th user in the estimate $\hat{x}_u$ and the term $\sum_{i=1,i \neq u}^{U} | \mathbf{f}_{\mathrm{BB},u}^H\mathbf{H}_{\mathrm{eff},u}^H \mathbf{H}_{\mathrm{eff},i} \mathbf{f}_{\mathrm{BB},i}|^2$ denotes the interference part. In the cooperative uplink scenario, the objective is to design the uplink hybrid TPCs for ensuring that they maximize the SINR of each user in the system. Hence, the pertinent optimization problem is given by
\begin{align}
& \underset{\mathbf{f}_{\mathrm{BB},u}}{\mathrm{maximize}}\quad \mathbf{f}_{\mathrm{BB},u}^H \left(\mathbf{H}_{\mathrm{eff},u}^H \mathbf{H}_{\mathrm{eff},u}\right) \mathbf{f}_{\mathrm{BB},u} \nonumber \\ & \text{s.t.}\quad\ \left(\mathbf{f}_{\mathrm{BB},i}^H \mathbf{H}_{\mathrm{eff},i}^H \mathbf{H}_{\mathrm{eff},u}\right) \mathbf{f}_{\mathrm{BB},u} =\mbox{0},\ i = \mbox{1},\mbox{2},\ldots, u-\mbox{1}, \nonumber \\ &\qquad\quad \|\mathbf{W}_{\rm RF}\mathbf{w}_{\mathrm{BB},u} \|_2^2 \leq P.
\label{UplinkOpt}
\end{align} 
From the first constraint in \eqref{UplinkOpt}, it can be readily observed that if the baseband TPC $\mathbf{f}_{\mathrm{BB},u}$ is designed so that any user $i$, $i < u$, does not face interference due to transmission of user $u$, then the same holds vice versa, i.e., the transmission of user $i$ does not create any interference for user $u$, $u > i$. Hence, one can deduce that the interference cancellation process is symmetric in nature. Furthermore, it can also be deduced that in the successive CFMU mmWave uplink system there is no interference among the scheduled users, which is in contrast to the successive downlink scenario. The solution to the optimization problem in \eqref{UplinkOpt} can be achieved by solving the following problem:
\begin{align}
& \underset{\mathbf{b}_u}{\mathrm{maximize}}\quad \mathbf{b}_u^H \left[\left(\mathbf{M}_{u-1}^{\perp}\right)^H \mathbf{H}_{{\rm eff},u}^H \mathbf{H}_{{\rm eff},u} \mathbf{M}_{u-1}^{\perp} \right] \mathbf{b}_u\nonumber \\ & \ \ \ \ \ \ \text{s.t.} \ \ \qquad \| \mathbf{b}_u \|_2^2 \leq P,
\label{Eq 9}
\end{align}
where $\mathbf{b}_u \in \mathbb{C}^{(N_{\mathrm{RF},u}-u+1)\times 1}$, and $\mathbf{M}_{u-1}^{\perp} \in \mathbb{C}^{N_{\mathrm{RF},u} \times (N_{\mathrm{RF},u}-u+1)}$ denotes a basis for the null space of the matrix $\mathbf{M}_{u-1} \in \mathbb{C}^{(u-1) \times N_{\mathrm{RF},u}}$, which is given as
\begin{equation*}
\mathbf{M}_{u-1} = \left[\begin{array}{c}\mathbf{f}_{{\rm BB},1}^H \mathbf{H}_{{\rm eff},1}^H \\ \mathbf{f}_{{\rm BB},2}^H \mathbf{H}_{{\rm eff},2}^H \\ \vdots \\ \mathbf{f}_{{\rm BB},u-1}^H \mathbf{H}_{{\rm eff},u-1}^H\end{array} \right] \mathbf{H}_{{\rm eff},u},
\end{equation*}
and can be obtained from the SVD of $\mathbf{M}_{u-1}$. Therefore, the uplink baseband TPC designed for the $u$th user is given as $\mathbf{f}_{\mathrm{BB},u} = \sqrt{P}\mathbf{M}_{u-1}^{\perp}\mathbf{b}_u^{*}$, where $\mathbf{b}_u^{*}$ denotes the dominant eigenvector of the matrix $(\mathbf{M}_{u-1}^{\perp})^H \mathbf{H}_{{\rm eff},u}^H \mathbf{H}_{{\rm eff},u} \mathbf{M}_{u-1}^{\perp}$. Furthermore, the rate $R_u$ of the $u$th user is given as $R_u = \log_2 (\mbox{1} + {\| \mathbf{H}_{{\rm eff},u} \mathbf{M}_{u-1}^{\perp} \mathbf{b}_{u}^{*} \|}_2^2)$, and the capacity $C_{\mathrm{sum}}$ in the uplink scenario is given as $C_{\mathrm{sum}} = \sum_{u=1}^{U}{R_u}$. Hence, one can notice that the proposed SCUHBF scheme maximizes each user's SINR, while successfully cancelling the IUI.
\begin{figure*}[t]
\centering
\subfloat[]{\includegraphics[width=0.45\linewidth]{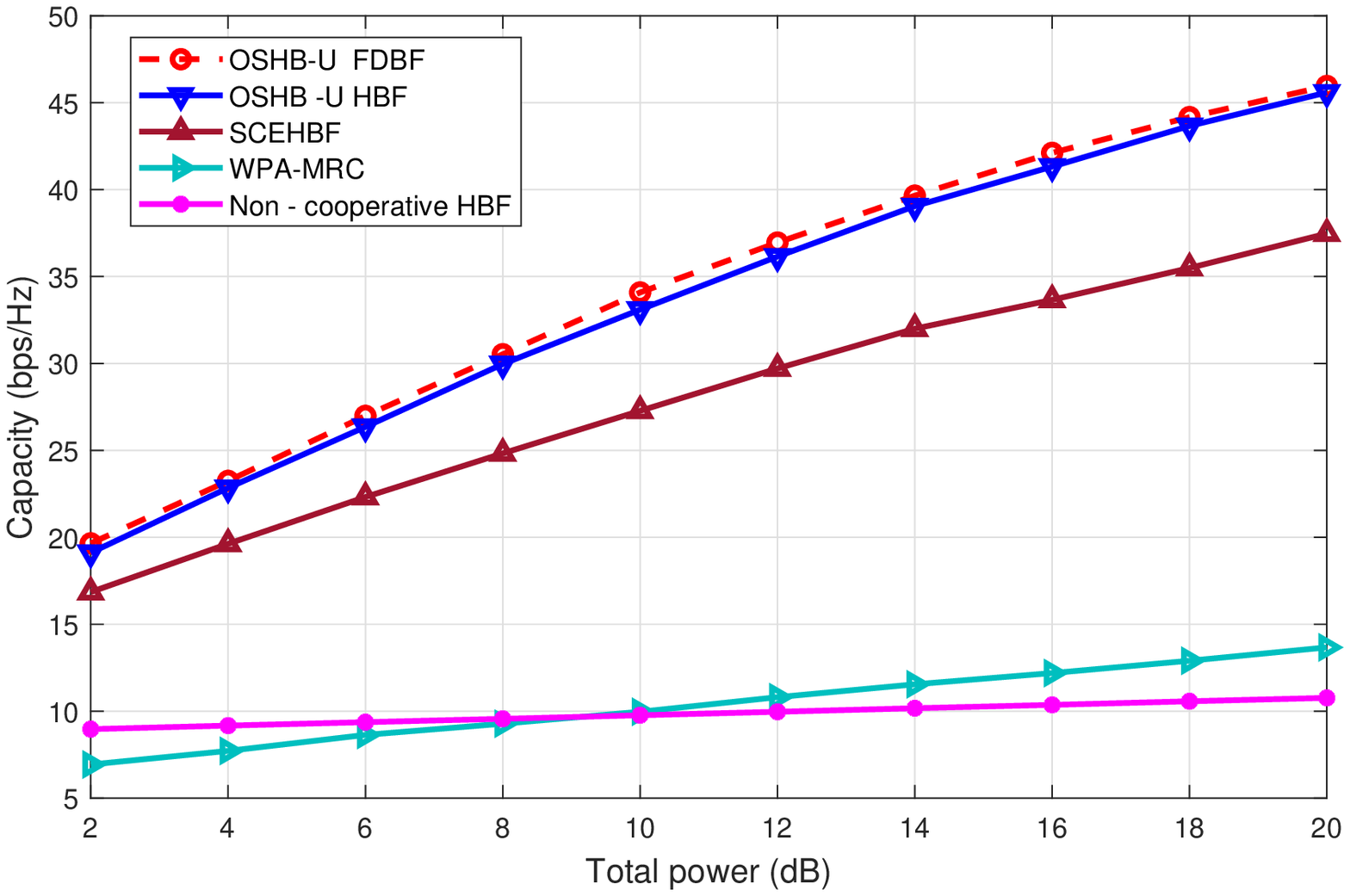}\label{4Ua}}
\subfloat[]{\includegraphics[width=0.45\linewidth]{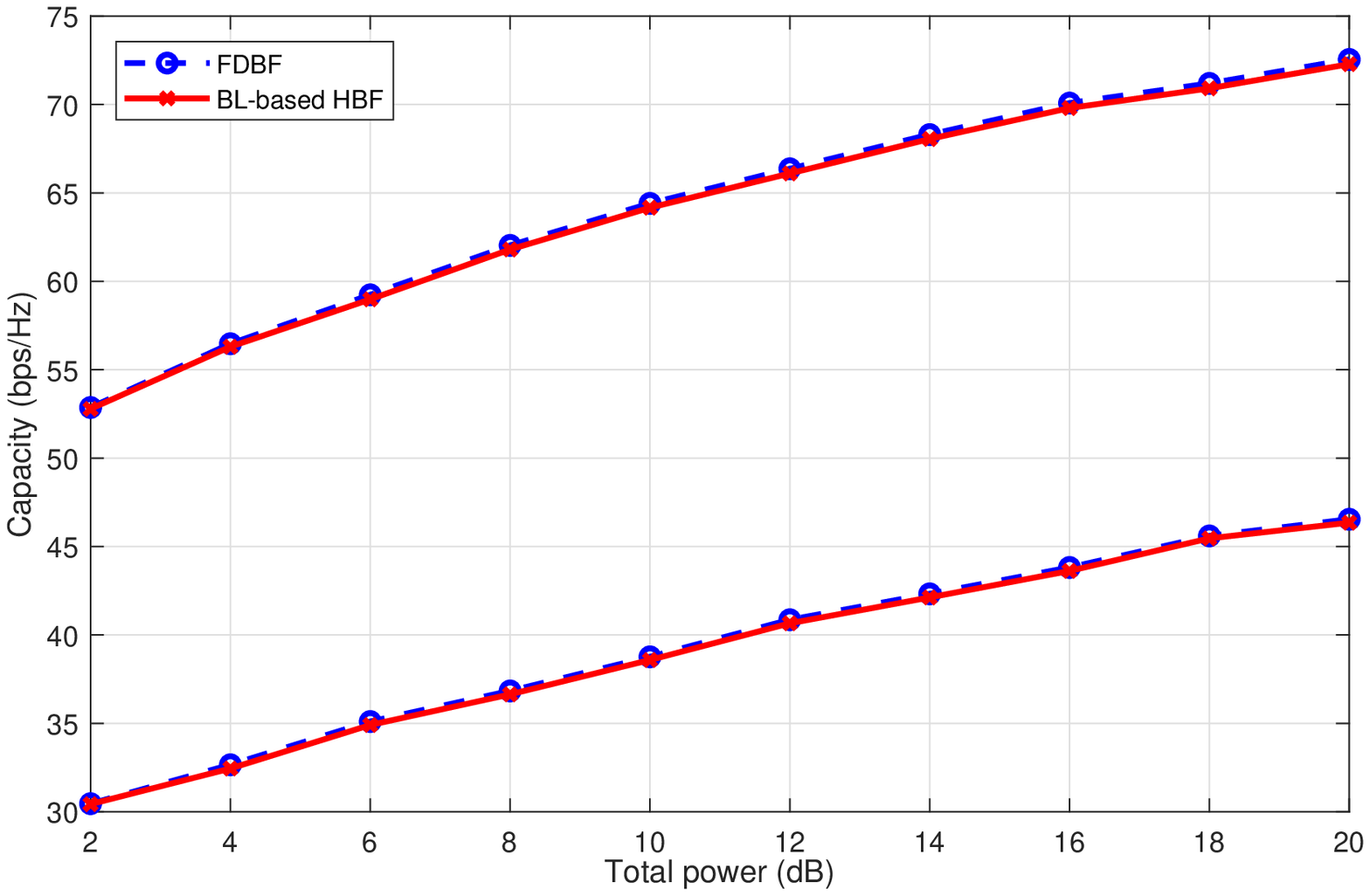}\label{4Ub}}
\caption{Unicast Scenario: (a) Capacity performance of the OSHB-U with various schemes for $N_T = 16$, $N_R = 8$, $N_{\mathrm{RF},m} = 4$, and $U = 4$, (b) capacity comparison of the fully-digital design and BL-based hybrid beamformer design for $N_T = 32$ and $N_R = 8$ antennas.\vspace{-15pt}}\label{UnicastFig}
\end{figure*}
\section{Simulation Results}
In this section, we provide simulation results for illustrating the performance of the proposed HTBFs in various scenarios. In the cell-free multi-user network considered, the $M$ APs and $U$ users are considered to be uniformly distributed at random within a square coverage area of size $200 \times 200$ $m^2$. The mmWave MIMO channel $\mathbf{H}_{u,m}$ between each user and the AP has been modelled using \eqref{Channel} with $L = 6$ multipath components. The AoA/AoDs of the channels between the APs and users are generated with uniform distribution over $\left[0 \ \ 2\pi\right]$. Furthermore, the multipath gains $\alpha_{u,m}^l$ are assumed to be complex Gaussian distributed as $\mathcal{C}\mathcal{N}\left(0,1\right)$. For the BL-based precoder and combiner design, the initial value of all the hyper-parameters is set to $\hat{\gamma}_i^{(0)} = 1, \forall 1 \leq i \leq S$, with the maximum number of EM iterations and the stopping parameter set as $k_{\mathrm{max}} = 50$ and $\epsilon = 10^{-6}$, respectively. We consider $3000$ such channel realizations for plotting the capacity $R_{\mathrm{sum}}$ versus the total power $P_T$ (dB) available at the APs. \vspace{-10pt}
\subsection{Beamforming for broadcast scenarios: HBBF}
For a broadcast scenario, a $\left(16 \times 8\right)$ elements mmWave MIMO system is considered, where we have $N_T = 16$ TAs at AP and $N_R = 8$ RAs at user, $M = 4$ APs and $U = 8$ users. For HTBF, we assume that the number of RF chains at each AP is equal to the number of users i.e. $N_{\mathrm{RF},m} = U$, and number of RF chains at each user is set equal to the total number of cooperative APs, i.e. $N_{\mathrm{RF},m} = M$. Figure \ref{Broadcast} illustrates the performance of the HBBF and HBBF-P schemes described in Section \ref{HBB}. It can be observed that the hybrid design of the proposed HBBF schemes closely approaches the performance of an ideal fully-digital beamformer (FDBF), where the number of RF chains is equal to the number of antennas. Furthermore, it can also be observed that the fairness-based HBBF-P scheme associated with per-AP power constraints closely follows the performance of the HBBF, given a pooled power constraint. Thus, the proposed HBBF-aided hybrid TPC design results in the optimal solution, and it is also well-suited for transmission in broadcast CFMU  cooperative mmWave MIMO scenarios. Fig. \ref{BroadcastLargeAntenna} illustrates the performance of the HBBF scheme in a large-scale antenna array regime. It can be observed that the capacity increases upon increasing the number of TAs at the AP, thanks to the array gain of the large antenna arrays. This further establishes the importance of having a large number of antennas in the mmWave regime for improved power efficiency.\vspace{-10pt}
%......................................................
\begin{figure*}
\centering
\subfloat[]{\includegraphics[height=5cm, width=8cm]{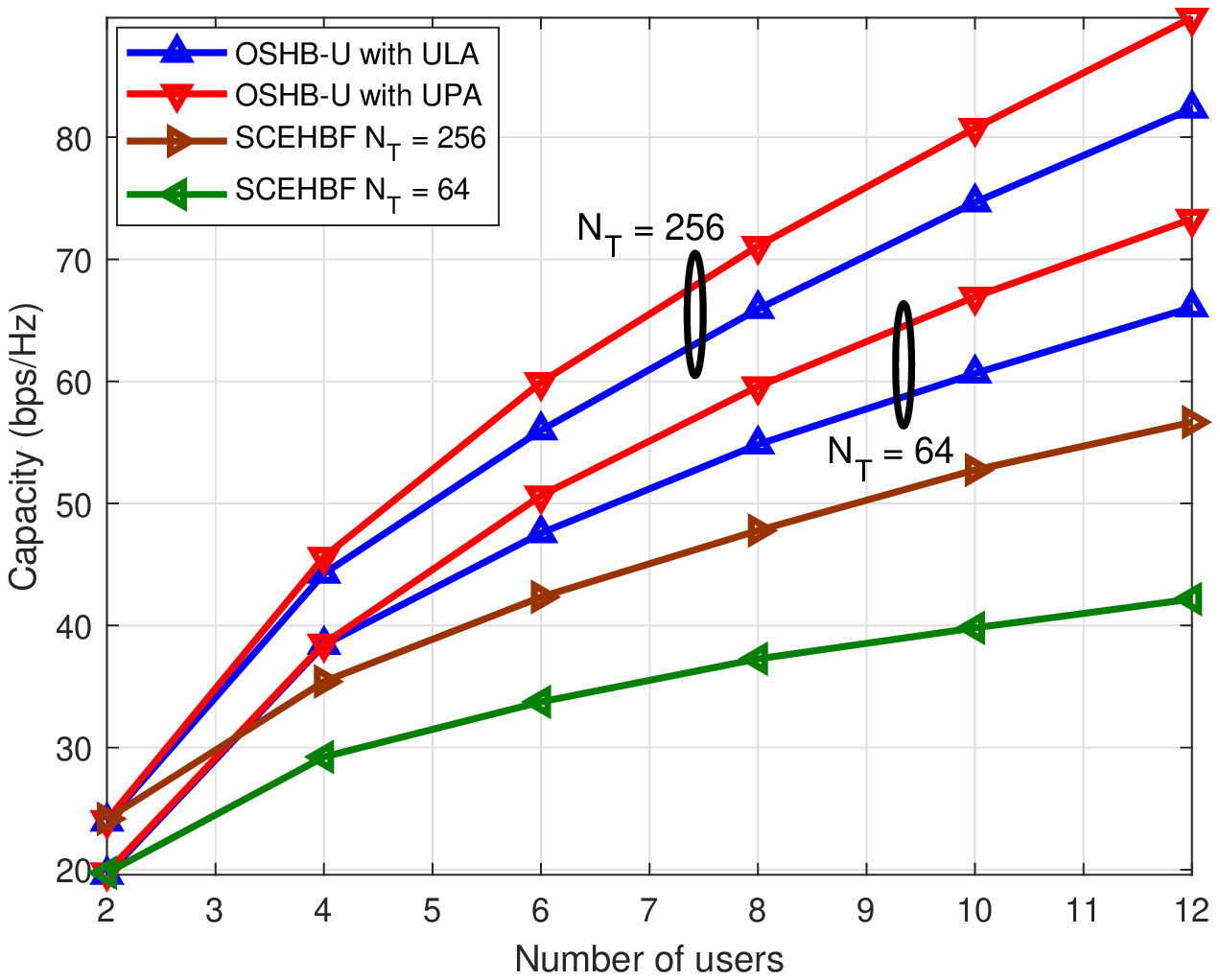}\label{UPA_ULAFig}}
\subfloat[]{\includegraphics[height=5cm, width=8cm]{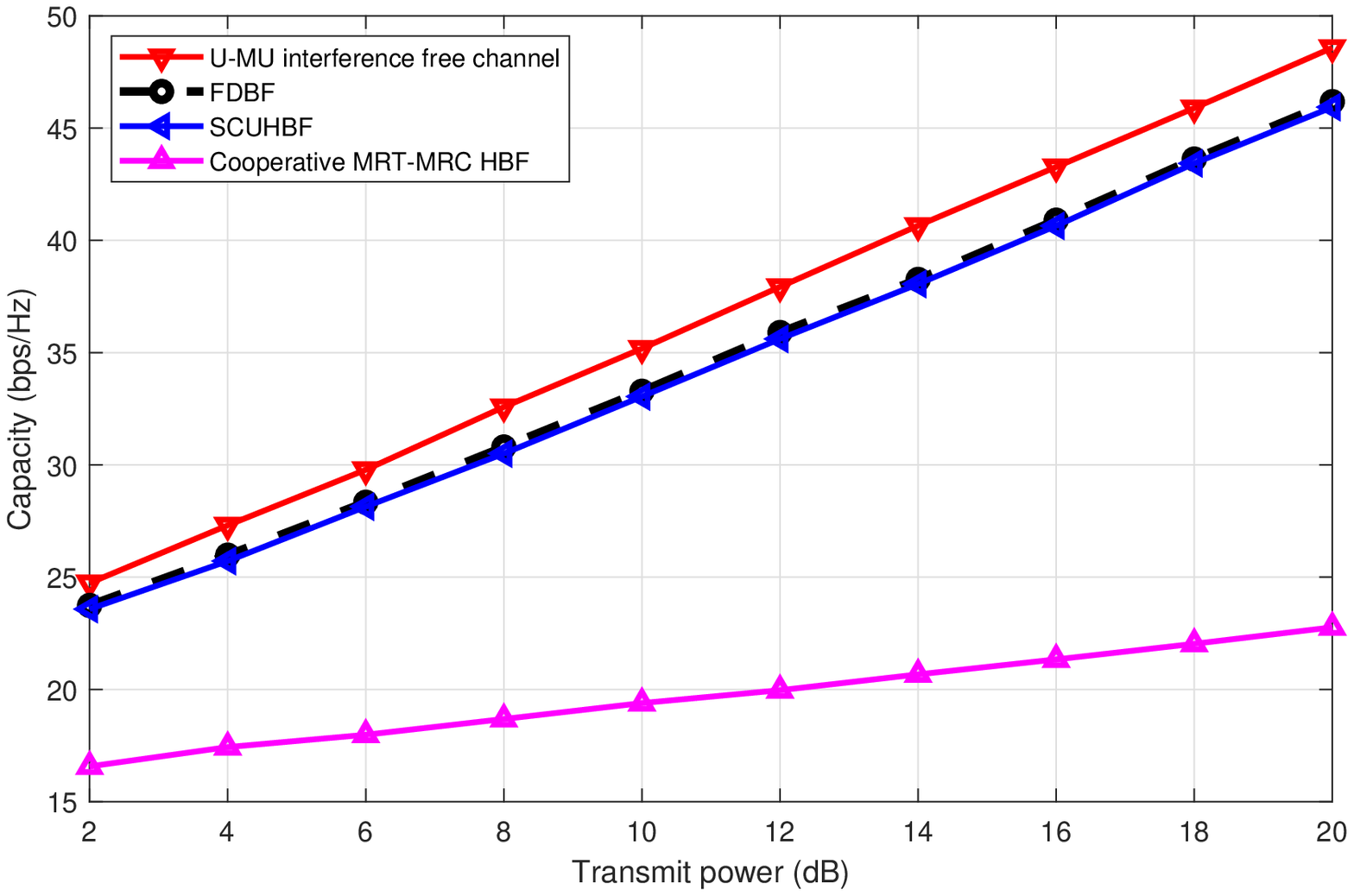}\label{UplinkFig}}
\caption{Unicast scenario: (a) Capacity versus number of users with $M = 4$, (b) Uplink scenario: capacity versus total power with $M = 4$, $N_T = 16$, $N_R = 8$, $G = 2$ and $U = 4$. \vspace{-10pt}}
\end{figure*}
%....................................................
\begin{figure*}[t]
\centering
\subfloat[]{\includegraphics[width=0.45\linewidth]{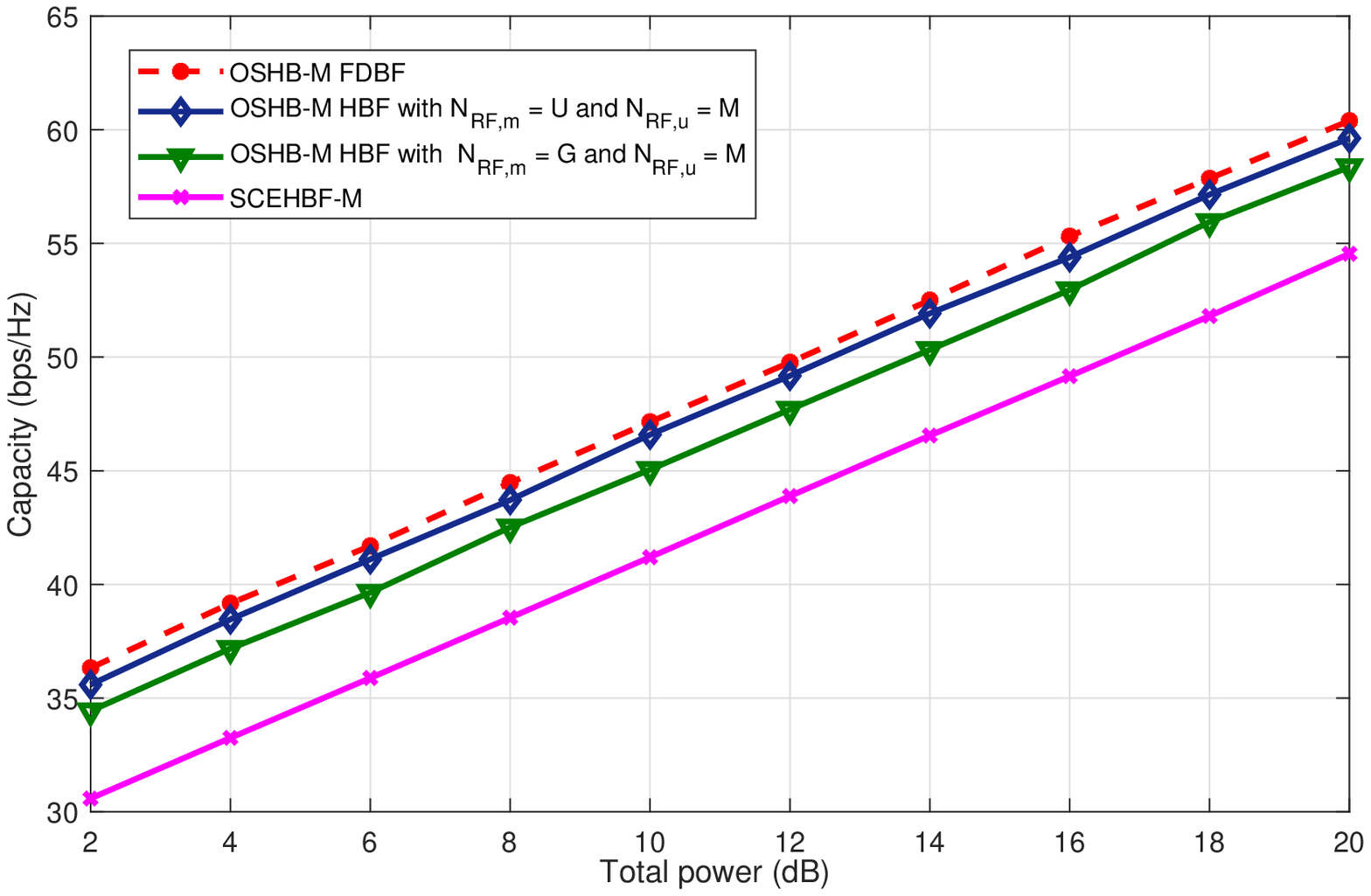}\label{MulticastFig}}
\subfloat[]{\includegraphics[width=0.45\linewidth]{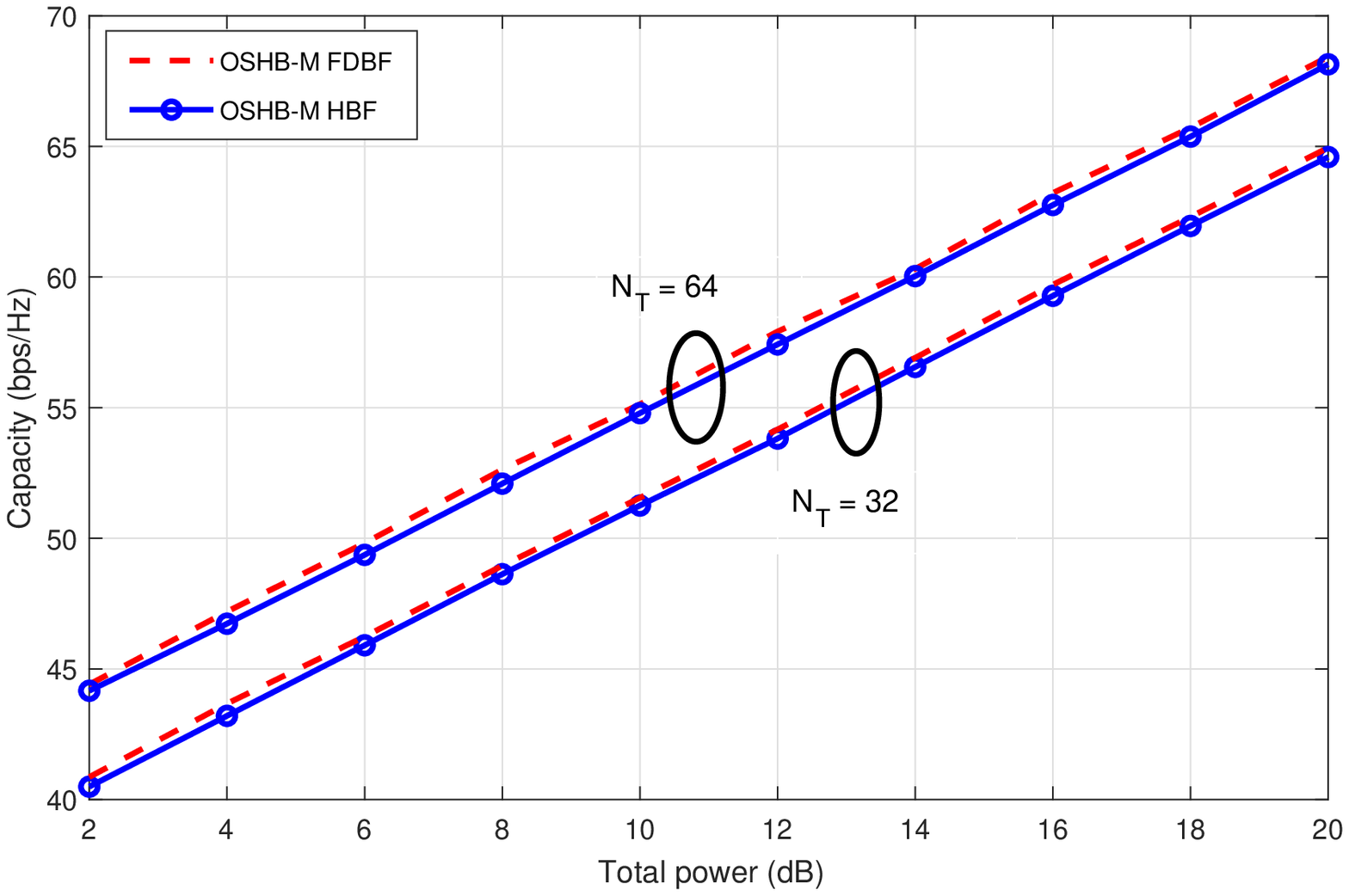}\label{MulticastLargeAntenna}}
\caption{Unicast Scenario: (a) Capacity versus total power for a multicast scenario with $M = 4$, $N_T = 16$, $N_R = 8$, $G = 2$ and $U_g = 4$, (b) capacity versus total power for a multicast scenario $M = 4$, $N_R = 8$, $G = 2$ and $U_g = 4$. \vspace{-10pt}}
\end{figure*}
\subsection{Beamforming schemes for downlink and uplink unicast scenarios: OSHB-U and SCUHBF}
Next, we consider both downlink and uplink CFMU mmWave MIMO unicast scenarios in the face of interuser interference and compare the capacity performance of various schemes. Fig. \ref{4Ua} shows the capacity performance of the OSHB-U in a downlink system along with those of the other schemes for $M = 4$ APs and $U = 4$ users with $N_T = 16$, $N_R = 8$, $N_{\mathrm{RF},m} = 4$ and $N_{\mathrm{RF},u} = M$. It can be observed that the OSHB-U scheme performs similarly to the ideal fully-digital beamformer. The proposed OSHB-U scheme achieves an improved performance in terms of capacity as a benefit of the SIC and the maximization of the SINR by the MVDR-based beamformer. For multi-cell cooperative scenarios, one can also employ the sub-optimal successive constrained eigen hybrid beamforming (SCEHBF) \cite{6607449} scheme based on MRC, which was originally derived for a sub-6 GHz system and has been suitably extended for mmWave for fair comparison. Its performance is poor, since it disregards the interference inflicted by the previously scheduled users. Furthermore, the simplistic white power allocation (WPA) \cite{6607449} scheme using MRC results in a significant performance degradation because it fails to cancel the interference.

Fig. \ref{4Ub} compares the capacity of the BL-based HTBF design with that of the ideal fully-digital beamformer. The BL-based HBF design does not require the full knowledge of AoAs/AoDs and their path gains for designing the baseband and RF precoders. It can be observed that the BL-based design attains a performance close to that of the fully-digital design with fewer RF chains. This is attributed to the fact that the mmWave MIMO channel is comprised of fewer multipath components, hence resulting in a low-rank channel. Therefore, the ideal FDBF can be approximated using a linear combination of only a few transmit array response vectors.

Fig. \ref{UPA_ULAFig} plots the capacity versus the number of users considering both the uniform planar array (UPA) and the uniform linear array (ULA) structures. Interestingly, it can be seen that the capacity increases upon increasing the number of AP antennas, which can be attributed to the fact that the CFMU mmWave MIMO system and the proposed design technique is capable of efficiently exploiting the resultant increased array gain. Furthermore, one can also observe that the proposed OSHBF-U algorithm outperforms the SCEHBF, when serving different number of users. To be specific, as the number of served users increases, the gap between the proposed and benchmark algorithms progressively increases, which further confirms the stability and effectiveness of our proposed algorithm.
Fig. \ref{UplinkFig} plots the capacity versus transmit power for the uplink scenario. For this analysis, $U=4$ active users are equipped with $N_R = 8$ TAs, and assumed to be distributed in $M = 4$ cooperating cells with each AP having $N_T = 16$ RAs.  It can be observed that the proposed SCUHBF scheme performs close to the capacity of $U$ interference-free point-to-point mmWave MIMO systems having dimensions of $\sum_{m=}^{M}N_{\mathrm{RF},m} \times N_{\mathrm{RF},u}$, and it also closely matches the performance of the ideal FD design. Furthermore, it is also observed that the capacity achieved by the proposed SCUHBF scheme for a cooperative uplink scenario is much higher than that of employing maximum ratio transmit (MRT) beamforming at the users and MRC at the APs.
\subsection{Beamforming for multicast scenarios: OSHB-M}
Fig. \ref{MulticastFig} compares the capacity performance of the OSHB-M scheme in the face of inter-group interference for different values of the number of RF chains, $M = 4$ APs and $U_g = 4$ users having $N_T = 16$, $N_R = 8$ antennas, and $G = 2$ user groups. It can be observed that the performance of the proposed OSHB-M scheme closely follows that of the ideal fully-digital TBF design, but with a much lower number of RF chains at the APs. Next, Fig. \ref{MulticastLargeAntenna} demonstrates the large-scale antenna benefits of the mmWave regime. It can be observed that the performance of the OSHB-M schemes improve significantly upon increasing the number of antennas at the APs. \vspace{-10pt}
%...........................
\section{Conclusion \vspace{-8pt}}
Optimal hybrid beamformers for broadcast, unicast and multicast CFMU mmWave MIMO systems are designed in the face of interuser/intergroup interference. Initially, for a broadcast scenario, the optimal beamforming solutions were derived by considering a total power constraint as well as per-AP power constraints. Subsequently, OSHB-based techniques relying on the optimal MVDR principle were proposed for unicast and multicast scenarios, which efficiently cancel the resultant interuser and intergroup interference. Next, BL-based joint hybrid transceiver was also design which did not require the knowledge of the AoAs/AoDs and the gains of the various multipath components of the mmWave MIMO channels. Finally, we also derived the SCUHBF-based cooperative uplink HBF, which maximizes the SINR, hence achieving high data rates. Our simulation results demonstrated the improved spectral efficiency as well as interference cancellation capability of the proposed HBBF, OSHB and SCUHBF techniques. The proposed hybrid designs have a low-complexity, and can simultaneously support an increased number of users in comparison to the existing methods, which makes them eminently suitable for practical mmWave CFMU MIMO systems. \vspace{-10pt}
\begin{appendices}
\section{MVDR Beamformer}
Considering the system model of \eqref{eq 16}, the signal $\tilde{\mathbf{y}}_u$ received at user $u$ after RF combining can be written as
\begin{align*}
\tilde{\mathbf{y}}_u = \mathbf{H}_{\mathrm{eff},u} \mathbf{M}_{u-1}^{\perp} \boldsymbol{\nu}_{u} x_u + \boldsymbol{\tau}_u,
\end{align*}
where $\boldsymbol{\tau}_u$ represents the interference-plus-noise component associated with the covariance matrix $\mathbf{C}_{\boldsymbol{\tau}_u}$. After baseband combining at user $u$, the signal estimate $\hat{x}_u$ is given as
\begin{align*}
\hat{x}_u = \mathbf{w}_{\text{BB},u}^H \mathbf{H}_{\mathrm{eff},u} \mathbf{M}_{u-1}^{\perp} \boldsymbol{\nu}_{u} x_u + \mathbf{w}_{\text{BB},u}^H \boldsymbol{\tau}_u.
\end{align*}
Therefore, the SINR at user $u$ is given by
\begin{align*}
\mathrm{SINR}_u = \frac{\vert \mathbf{w}_{\text{BB},u}^H \mathbf{H}_{\mathrm{eff},u} \mathbf{M}_{u-1}^{\perp} \boldsymbol{\nu}_{u} \vert^2}{\mathbf{w}_{\text{BB},u}^H \mathbf{C}_{\boldsymbol{\tau}_u}\mathbf{w}_{\text{BB},u}}.
\end{align*}
The MVDR beamformer is the optimal beamformer that maximizes the SINR \cite{van2002modulation}, and it is given as $\mathbf{w}_{\text{BB},u} = \mathbf{C}^{-1}_{\boldsymbol{\tau}_u} \mathbf{H}_{\mathrm{eff},u} \mathbf{M}_{u-1}^{\perp} \boldsymbol{\nu}_{u}$. Furthermore, the maximum value of the SINR is determined as

\begin{align*}
\mathrm{SINR}_{\mathrm{max},u}=&\, \frac{\vert \boldsymbol{\nu}_{u}^H \mathbf{M}_{u-1}^{\perp H} \mathbf{H}_{\mathrm{eff},u}^H \mathbf{C}^{-1}_{\boldsymbol{\tau}_u} \mathbf{H}_{\mathrm{eff},u} \mathbf{M}_{u-1}^{\perp} \boldsymbol{\nu}_{u} \vert ^2}{\boldsymbol{\nu}_{u}^H \mathbf{M}_{u-1}^{\perp H} \mathbf{H}_{\mathrm{eff},u}^H \mathbf{C}^{-1}_{\boldsymbol{\tau}_u} \mathbf{H}_{\mathrm{eff},u} \mathbf{M}_{u-1}^{\perp} \boldsymbol{\nu}_{u}} \\ =&\, \boldsymbol{\nu}_{u}^H \mathbf{M}_{k-1}^{\perp H} \mathbf{H}_{\mathrm{eff},u}^H \mathbf{C}^{-1}_{\boldsymbol{\tau}_u} \mathbf{H}_{\mathrm{eff},u} \mathbf{M}_{u-1}^{\perp} \boldsymbol{\nu}_{u}.
\end{align*}
Observe from the above expression that the optimal vector $\boldsymbol{\nu}_{u}$ that maximizes the $\mathrm{SINR}_{\mathrm{max}, u}$ is the principal right singular vector of $\mathbf{C}^{-(1/2)}_{\boldsymbol{\tau}_u} \mathbf{H}_{\mathrm{eff},u} \mathbf{M}_{u-1}^{\perp}$. Therefore, the optimal MVDR receive beamformer $\mathbf{w}_{\mathrm{BB},u}$ is given by $\mathbf{w}_{\mathrm{BB},u} = {(\mathbf{C}_{\boldsymbol{\tau}_{u}}^{-(1/2)})}^H \tilde{\mathbf{w}}_{\mathrm{BB},u}$, where $\tilde{\mathbf{w}}_{\mathrm{BB},u}$ is the dominant left singular vector
of the matrix $\mathbf{C}^{-(1/2)}_{\boldsymbol{\tau}_u} \mathbf{H}_{\mathrm{eff},u} \mathbf{M}_{u-1}^{\perp}$. This completes the proof. \vspace{-12pt}
\end{appendices}
\bibliographystyle{ieeetran}
\bibliography{biblo}
\end{document}